\documentclass[sigconf]{acmart}

\AtBeginDocument{%
  \providecommand\BibTeX{{%
    \normalfont B\kern-0.5em{\scshape i\kern-0.25em b}\kern-0.8em\TeX}}}

\usepackage{subcaption}

\begin{document}

\title[\system{}]{\system{}: Optimization-based Mixed Reality Workspace Integration for Remote Side-by-side Collaboration}

\author{Ludwig Sidenmark}
\orcid{0000-0002-7965-0107}
\affiliation{%
  \institution{University of Toronto}
  \city{Toronto}
  \state{Ontario}
  \country{Canada}
}
\email{lsidenmark@dgp.toronto.edu}

\author{Tianyu Zhang}
\orcid{0009-0008-8035-8040}
\affiliation{%
  \institution{University of Toronto}
  \city{Toronto}
  \state{Ontario}
  \country{Canada}
}
\email{tianyuz@dgp.toronto.edu}

\author{Leen Al Lababidi}
\orcid{0009-0001-6331-8097}
\affiliation{%
  \institution{University of Toronto}
  \city{Toronto}
  \state{Ontario}
  \country{Canada}
}
\email{l.lababidi@mail.utoronto.ca}

\author{Jiannan Li}
\orcid{0000-0001-8409-4910}
\affiliation{%
  \institution{Singapore Management University}
  \city{Singapore}
  \country{Singapore}
}
\email{jiannanli@smu.edu.sg}

\author{Tovi Grossman}
\orcid{0000-0002-0494-5373}
\affiliation{%
  \institution{University of Toronto}
  \city{Toronto}
  \state{Ontario}
  \country{Canada}
}
\email{tovi@dgp.toronto.edu}

\newcommand{\system}[0]{Desk2Desk}
\newcommand{\hl}[1]{{\color{black}#1}}

\copyrightyear{2024} 
\acmYear{2024} 
\setcopyright{acmlicensed}\acmConference[UIST '24]{The 37th Annual ACM Symposium on User Interface Software and Technology}{October 13--16, 2024}{Pittsburgh, PA, USA}
\acmBooktitle{The 37th Annual ACM Symposium on User Interface Software and Technology (UIST '24), October 13--16, 2024, Pittsburgh, PA, USA}
\acmDOI{10.1145/3654777.3676339}
\acmISBN{979-8-4007-0628-8/24/10}


\begin{abstract}
Mixed Reality enables hybrid workspaces where physical and virtual monitors are adaptively created and moved to suit the current environment and needs. However, in shared settings, individual users’ workspaces are rarely aligned and can vary significantly in the number of monitors, available physical space, and workspace layout, creating inconsistencies between workspaces which may cause confusion and reduce collaboration. We present \system{}, an optimization-based approach for remote collaboration in which the hybrid workspaces of two collaborators are fully integrated to enable immersive side-by-side collaboration. The optimization adjusts each user’s workspace in layout and number of shared monitors and creates a mapping between workspaces to handle inconsistencies between workspaces due to physical constraints (e.g. physical monitors). We show in a user study how our system adaptively merges dissimilar physical workspaces to enable immersive side-by-side collaboration, and demonstrate how an optimization-based approach can effectively address dissimilar physical layouts.
\end{abstract}




\begin{teaserfigure}
  \centering
  \includegraphics[width=1.0\linewidth]{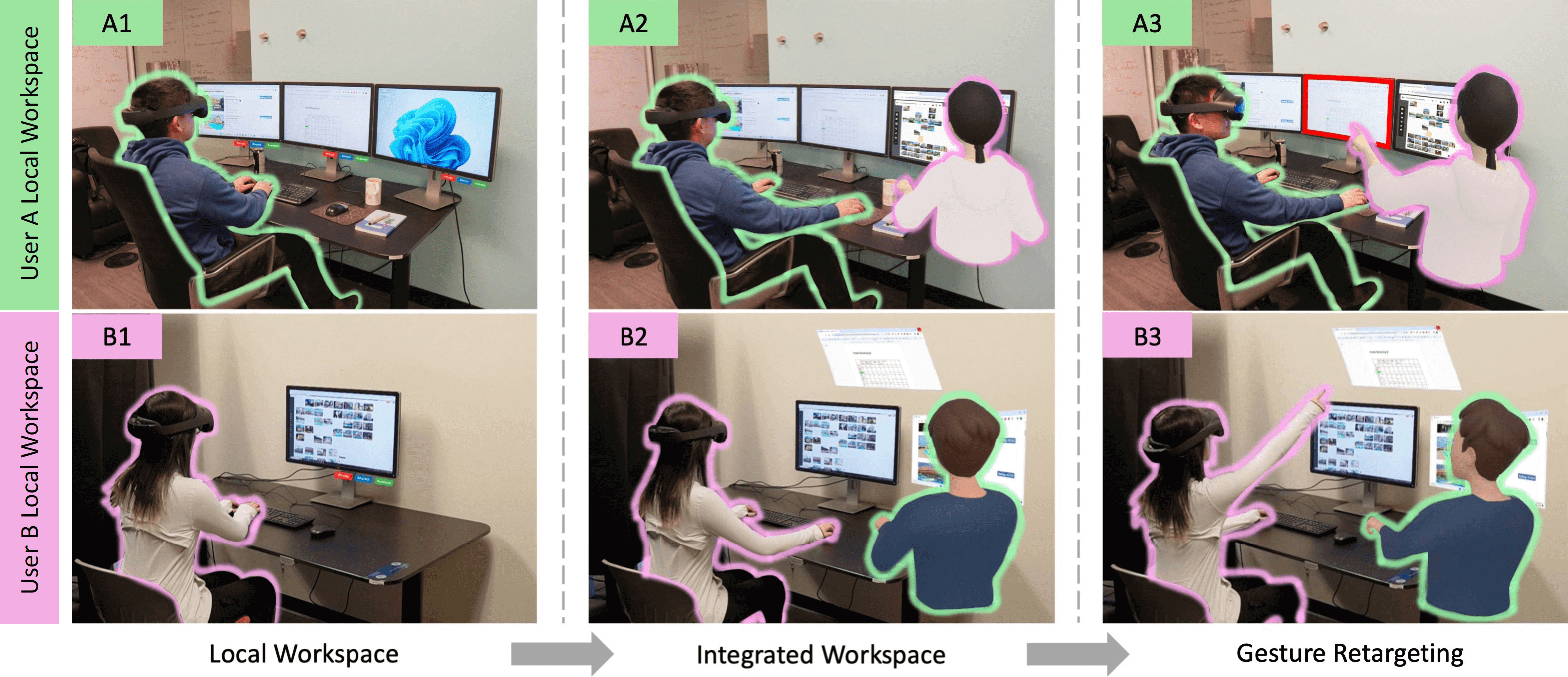}
  \caption{In \system{}, remote collaborators meet in a distributed Mixed Reality space composed of their local physical workspace for side-by-side collaboration. Users define their workspace by selecting the screens to share and screens available for sharing to (A1 and B1). After connecting, integration is performed via optimization objectives to unify the varying workspaces. \system{} creates and places virtual screens in optimal positions (B2) or overlays them on existing available screens (A2) as needed. During collaboration, local users' physical interaction, e.g. pointing at screens (B3), or gazing toward the collaborator (A3) is kept consistent by retargeting the remote avatars to ensure spatial consistency between dissimilar workspaces.}
  \Description{This figure illustrates two users, A and B, transitioning from a local to a virtual workspace using VR headsets. For User A and User B, the sequence shows:
Local Workspace - working with physical monitors, keyboard, and mouse.
Integrated Workspace - interacting with a mix of physical and virtual screens using hand gestures. Gesture Retargeting - using gestures to interact with virtual avatars, indicating advanced virtual collaboration. Both rows demonstrate the merging of real and virtual work environments, highlighted by silhouettes outlined in green and pink.}
  \label{fig:teaser}
\end{teaserfigure}

\maketitle

\section{Introduction} 

Extended Reality (XR) allows users to meet remotely and supports visual aids, shared visual attention, and gestures beyond traditional physical desktop screens to enable tightly connected and immersive collaboration~\cite{Bovo2023Cov, Bai2020Collaboration, Allison2021EyeMrVis}. This has sparked a number of works where remote or co-located collaborators meet in fully virtual environments ~\cite{Bovo2023Cov, Frohler2022Xva}, or in Mixed Reality (MR) spaces where physical and virtual objects are merged and aligned between collaborators to integrate physical objects into the environment~\cite{Gronbaek23, Frohler2022Xva}. Thus far, such systems have mainly focused on collaborative scenarios where users are surrounded by virtual screens~\cite{Frohler2022Xva}, or multiple separated physical surfaces, which they move between to perform tasks~\cite{Gronbaek23}. In contrast, most knowledge work (i.e., programming or document writing) is performed in a sitting posture in front of a physical workspace with one or multiple screens. Remote collaboration is then typically performed through videoconferencing tools such as Zoom\footnote{https://zoom.us/}, or collaborative applications such as Figma\footnote{https://www.figma.com/}. However, such desktop-based tools lack the immersion and attention cues provided by physical collaborators, and there is also a lack of immersive MR work for workspace-based remote collaboration. In this work, we seek more immersive solutions for working with a remote partner in a seated side-by-side fashion, such as when performing pair programming or co-writing.

Integrating virtual elements into a physical environment is challenging due to the varying physical limitations of the environment. This has spurred multiple works on \emph{layout optimization} that attempt to find the most optimal position for virtual elements based on set criteria~\cite{Cheng21SemanticAdapt, Cheng2023InteractionAdapt}. This aspect becomes even more challenging for remote MR collaboration, as both users will be in completely separate physical environments, and any collaborative system must consider and unify these dissimilar spaces to avoid confusion and reduced collaboration caused by inconsistencies between environments~\cite{Gronbaek23}. Specifically for seated workspaces, the number of screens and their layout can vary significantly~\cite{yuan2022understanding}. This creates challenges for remote collaboration, as the deictic and physical cues normally present during local side-by-side collaboration can be potentially detrimental due to the dissimilar screen layouts.

In this work, we present \system{} (\autoref{fig:teaser}), an optimization-based collaborative system where two collaborators' MR workspaces consisting of both virtual and physical screens, and the virtual avatars of the two collaborators are aligned to allow side-by-side remote collaboration for desktop-based knowledge work. By building on \emph{blended environments}~\cite{Gronbaek23, Fidalgo23}, and \emph{layout optimization}~\cite{Cheng21SemanticAdapt, Niyazov23}, \system{} considers the positions of each workspace screen and user position, their spatial relationships together with environmental constraints to align the workspaces. Each user defines how many screens they wish to share to their collaborator (\autoref{fig:teaser}A1-B1). Our approach then computes the optimal layout for each user using integer programming to merge the workspaces. \system{} dynamically spawns and places virtual screens to handle any discrepancies between workspaces to support the number of screens each user wishes to share (\autoref{fig:teaser}A2-B2), and creates a mapping between linked screens to support a virtual representation of each collaborator that visualizes their attention through head and hand movements. Any discrepancy caused by physical limitations of the workspace is handled by a retargeting system to support head and hand gestures irrespective of workspace layout (\autoref{fig:teaser}A3-B3). \system{} enables side-by-side remote MR collaboration in front of a workspace irrespective of differences in physical environment and each user's initial workspace.

We evaluated \system{} in a collaborative remote task to study its ability to support remote MR side-by-side collaboration. The results of our study show that the system supports collaboration across a wide range of workspace combinations. Furthermore, results showed that \system{} supports physical hand and head gestures, increased perceived presence, usability, and a sense of understanding of each other's actions compared to a typical videoconference baseline. Our work showcases how optimization-based approaches used in layout optimization can be extended to collaborative settings for establishing common ground between remote collaborators in dissimilar physical environments. In sum, we provide the following contributions:
\begin{itemize}
    \item \system{}, a MR optimization-based system to integrate desktop workspaces for remote side-by-side collaboration.
    \item Results from a user study that show that \system{} can effectively support side-by-side collaboration across multiple workspace combinations, and encourages physical, desktop-based and verbal communication.
\end{itemize}
\section{Related Work} 

We build on prior work on XR collaboration, multi-screen workspaces, and adaptive interfaces.

\subsection{Shared Workspaces in XR} 

Computer-Supported Cooperative Work (CSCW) research pointed out that successful group work relies on maintaining common ground~\cite{clark1991grounding}, that is, shared knowledge with other participants that enables effective and efficient communication~\cite{kraut2003visual}.
Productive collaboration also requires adequate workspace awareness~\cite{gutwin2002descriptive}, including up-to-date understanding of task progress and collaborators' explicit and implicit communication cues. 
Research has further highlighted the importance of nonverbal cues, such as gaze, gestures, and body orientation, and deictic expressions (e.g. ``this'' and ``there'') for establishing and maintaining both common ground and workspace awareness. The use of these cues first helps users establish shared visual attention, and thereby conversational common ground~\cite{kirk2007turn,shockley2009conversation,burgoon2003nonverbal}.
Perhaps the most common example of such uses is performing a pointing gesture to refer to certain objects~\cite{Wong2014Gestures}. 
Such non-verbal cues also support people's awareness of the collaborators to infer their intentions and knowledge of the workspaces~\cite{Dangelo17GazeProgramming, Efran1968Collaboration, Nguyen2005MultiView, Schneider2013Collaboration, Vertegaal2003Gaze2}, or the desire to communicate~\cite{Schneider2013Collaboration, Dangelo17GazeProgramming}.

In contrast to traditional videoconferencing systems (i.e., Zoom), which lack a shared frame of reference to support non-verbal cues~\cite{Ohara2011Blended}, the extended tracking in XR allows significant potential for immersive and effective collaboration. Previous work has shown how presenting remote collaborators as avatars to enable natural hand and gaze gestures~\cite{Sousa19, Piumsomboon2018MiniMe, Hoppe2021Shisha}, or by adding visual cues that represent collaborators' gaze attention or pointing direction~\cite{Piumsomboon19, Bovo2023Cov, Allison2021EyeMrVis} improve collaboration, presence, and task performance. 

In addition, collaborative XR systems are not confined to a physical screen area~\cite{Biener2020BreakingTheScreen, Thoravi2019Loki}, and content can be presented in various configurations. This has led to extensive experimentation in which virtual content is placed around users in spherical~\cite{Satriadi2020Maps}, cylindrical~\cite{Bovo2023Cov}, or square formations~\cite{Lee2021Collaboration}. Alternatively, virtual content is attached to physical surfaces or objects to anchor collaboration in the physical environment~\cite{Gronbaek23}. The commonality with all these works is that the additional space for the content provided by XR has led to a focus on collaborative scenarios in which users are standing and actively moving between content to leverage the extra physical space enabled by XR~\cite{Lee2021Collaboration, Gronbaek23, Bovo2023Cov, Luo22Collaboration}. However, contemporary knowledge work is still performed primarily in desk workspaces in front of physical screens to allow work to be carried out throughout the day without physical fatigue~\cite{yuan2022understanding}. 
Collaborative tasks in such environments (e.g., pair programming or research~\cite{Williams2002Programming, Morris2008WebSearch} and collaborative web search~\cite{Morris2008WebSearch}) usually take place side-by-side in a seated position facing the workspace~\cite{Marquardt2012CrossDevice}. We build on knowledge from previous collaborative XR systems to design a system for seated remote collaboration in front of an MR workspace that integrates the screens of the user's workspace together with a virtual avatar of the remote collaborator and visual attention cues.

\subsection{Multi-Screen Workspaces}
To support the increasing demands of productivity work, physical workspaces have increased significantly in the number of devices and screens used~\cite{yuan2022understanding}.
\hl{Even in XR which promise vastly expanded workspaces, users still value physical displays mainly due to familiarity and superior resolution~\cite{Pavanatto21}.}
Multi-screen workspaces enable users to access more information~\cite{Endert2012Workspace}, increase productivity by enabling more efficient multi-tasking~\cite{Czerwinski2003Workspace}, and have proven to increase productivity in a wide variety of productivity contexts such as accounting~\cite{Collins2011Accounting}, engineering and office work~\cite{Ling2017Workspace, Owens2012Workspace}, and spatial tasks~\cite{Tan2006Workspace}. In multi-screen setups, screens are commonly categorized into primary screens used for the current main activity and interaction (i.e., programming IDE), while secondary screens are placed in the periphery to support the primary task (i.e., code documentation), screen communication, or for storing information such as to-do lists~\cite{Grudin2001MultiDisplay, Hutchings2004MultiDisplay}. Depending on user needs and spatial limitations, workspaces can be configured widely differently and with a wide variety in the number of screens~\cite{yuan2022understanding}. However, since physical screens require physical desk or wall space, effective configurations are not always available depending on the environment~\cite{Williamson2019PlaneVr, Medeiros2022PassengerWorkspace, Ng2021PassengerWorkspace}.

\subsection{Creating Coherent MR Environments} 
\hl{
MR workspaces extend beyond the constraints of physical environments and offer productivity benefits including significantly larger spaces~\cite{Biener2020BreakingTheScreen, Mcgill2020Workspaces}, effective document operations~\cite{Zhen2019HoloDoc, Rajaram2022PaperTrail, Luo22Collaboration, Qian2022DuallyNoted}, 
accommodating 3D data~\cite{Plasson2022DesktopAR}, and dynamic adaption to task needs~\cite{Pavanatto2021VirtualScreens, Medeiros2022PassengerWorkspace}. 
However, fitting virtual contents into physical environments can be difficult due to the wide variety of possible environmental constraints (i.e., walls). As manual placement could be tedious~\cite{Pavanatto2021VirtualScreens}, researchers have developed \emph{adaptive interface} models that place virtual content in the most optimal way based on a set of objectives and constraints~\cite{Lindlbauer2022Adaptive}. Previous work has shown how adaptive interfaces can effectively adapt to user workload and movements~\cite{Lindlbauer19, Belo2022Auit}, semantic connections between interface elements and physical objects~\cite{Cheng21SemanticAdapt, Qian2022Scalar}, user behavior over time~\cite{Fender2017Heatspace}, interaction affordances~\cite{Cheng2023InteractionAdapt}, and constraints defined by the user~\cite{Niyazov23}. Building on prior work on single-user layout optimization, we propose an method based on workspace semantics to align multiple physical environments and enable remote side-by-side collaboration.

Remote collaboration in MR could be more challenging than in VR as the physical environments in which the collaborators are located usually do not align and are difficult to modify~\cite{Wang2021Review}. Furthermore, collaborators in different reference frames may face difficulties when interpreting deictic references~\cite{Yoon21}, or referring to physical objects (e.g., tables) for interaction~\cite{Gronbaek23, Groenbaek2024Whiteboard}. Therefore, multiple works have attempted to align dissimilar physical environments to create a common frame of reference~\cite{Yoon21, Gronbaek23, Lehment2014Alignment}. However, full alignment can be challenging if the environments are not similar. 
\citet{Gronbaek23} proposed the concept of \emph{Partially Blended Environments} where only the physical objects necessary for collaboration (e.g., tables or whiteboards) are aligned to simplify the process. We take inspiration from this concept by only considering the screens for alignment to simplify the integration of workspaces. 

Another challenge for remote MR collaboration is the possible misalignment between remote collaborators' avatars when they are projected into the same physical environment~\cite{Pejsa2016Room}. For example, consider the scenario where both users are on the right side of the collaborator's avatar due to space restrictions. Whenever the user looks at the collaborator, from the collaborator's perspective the user would look away from them. To avoid such mismatch, collaborative systems employ \emph{retargeting} methods such as inverse kinematics to solve deictic gesture inconsistencies~\cite{Gronbaek23, Yoon21, Fink2022ReLocations}, to improve their accuracy~\cite{Sousa19},  or to share common views while retaining face-to-face positions to improve collaboration~\cite{Fidalgo23}. We leverage gesture and gaze retargeting to create a common frame of reference between remote users and workspaces, and use optimization algorithms to maximize the similarity between workspaces and minimize the need for retargeting.
}
\section{System} 

\system{} is designed for two collaborators to integrate their hybrid workspaces into one single shared workspace for side-by-side collaboration. This new merged layout takes into consideration differences and limitations in each user's physical environments and uses their original personal workspace as a basis. To enable this, \system{} consists of two key features: 1) integrating and optimizing the layout of the shared workspaces; and 2) maintaining the consistency of the avatar representation of the users during collaboration. To start an integration of workspaces, the users first label their screens to decide their role in the optimization:  

\begin{description}
\item [Private:] Screen that is hidden to the collaborator and is not considered in the optimization. 
\item [Shared:] Screen that should be shared to the collaborator.
\item [Available:] Screen that is unused by the local user and can be used to cast a shared screen to. 
\end{description}

\begin{figure}[t!]
    \centering
    \includegraphics[width=\columnwidth]{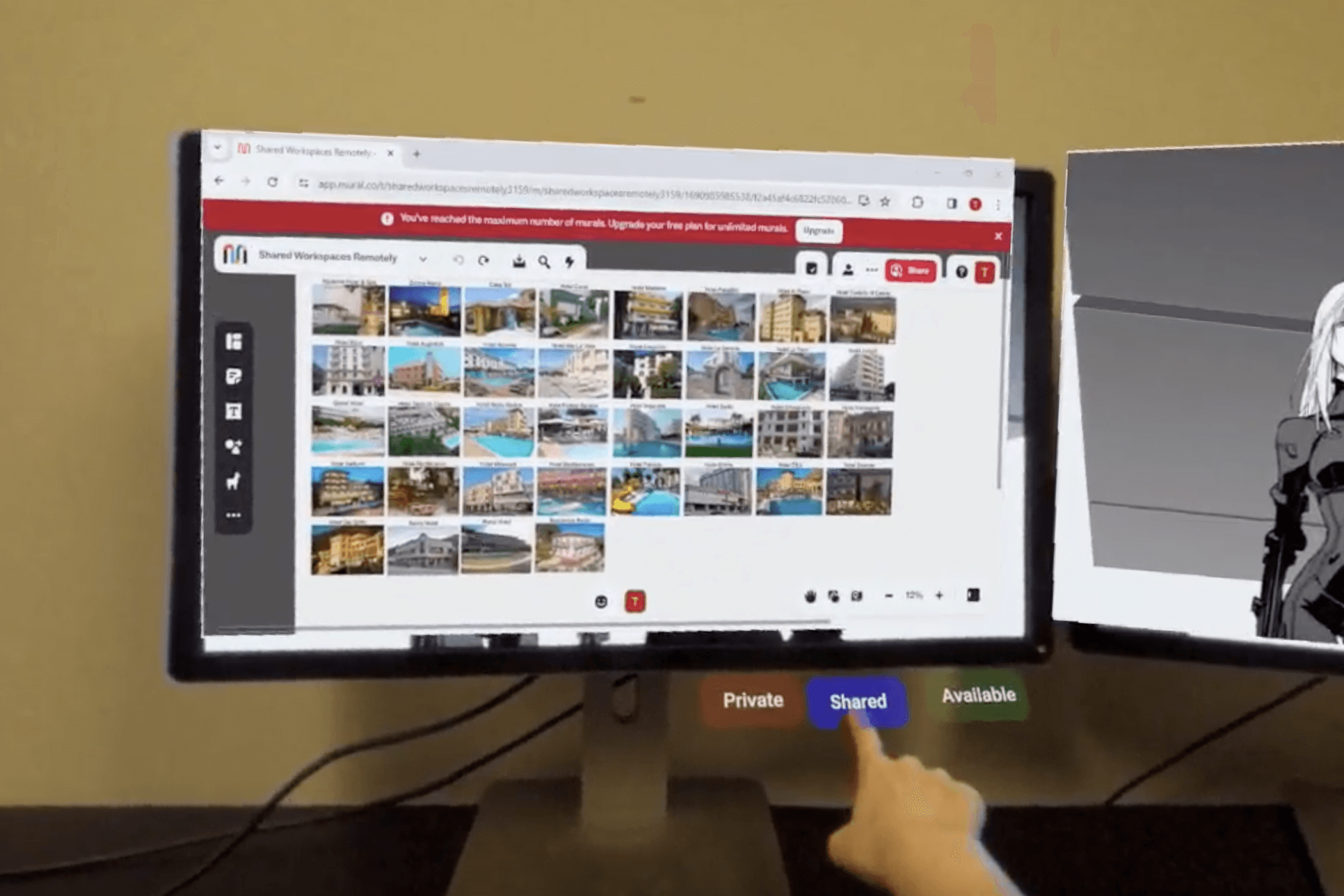}
    \caption{The user labeling a physical screen as \emph{Shared} through direct touch.}
    \Description{This image shows a monitor with a webpage full of image thumbnails, with a banner prompting an upgrade for unlimited access. Below are virtual buttons labeled "Private," "Shared," and "Available," with a finger pressing "Shared."}
    \label{fig:screen_label}
\end{figure}

Labeling is performed by direct touch input on virtual AR toggle buttons placed below the screen (\autoref{fig:screen_label}). Users then connect to their collaborator which starts the optimization. Screens are then generated and positioned based on the unique combination of physical and virtual screen from the users' workspaces. Finally, since workspaces are restricted by the physical constraints of their physical monitors and layout, the system creates a mapping between the screens in each user's view of the workspace, which is used to ensure gestural consistency during collaboration. 

\subsection{Virtualizing the Workspace}
The first challenge is to transform a hybrid workspace, consisting of physical and virtual monitors, into a virtual representation used for optimization and then displayed to the remote collaborator. The workspace and its physical surroundings are predefined by the local user by \hl{outlining} the positions $p_s$ and dimensions $d_s$ of physical screens, physical constraints (e.g., walls and tables), and the available space where virtual screens can be placed or spawned by the optimizer\hl{. This approach aligns with} previous work~\cite{Cheng2023InteractionAdapt, Cheng21SemanticAdapt} and gives users agency over their local space~\cite{Niyazov23}. All are outlined manually by the user by drawing bounding cubic volumes and labeling them. \hl{The user also defines the position of the remote collaborator during collaboration to ensure agency over their personal space relative to the collaborator.} We use the \emph{Oculus Room Capture} feature for this purpose and assume that each user has individually set up their local workspace environment prior to integration. However, recording of the local physical environment could be performed implicitly in real-time with computer vision or other sensing methods. After integration, the content of shared screens is streamed to the collaborator, similar to a typical videoconferencing screen sharing. As such, the local host is the only one who can manipulate the shared screen content using the standard keyboard and mouse. The remote collaborator can view the content on the shared screen, but is unable to directly interact with it.

\subsection{Optimization Scheme}
After the two collaborators have connected, the optimizer will attempt to find an optimal combined workspace for each user. The goal of the optimizer is to define an association between screens and potential locations in an environment, referred to as \emph{``containers''}. The optimizer uses the virtual representation of each user's workspace as input. Since the two users are in different physical environments, they will each have a unique set of environmental constraints. For example, in one case, a physical object (e.g., a wall or shelf) may block a potential screen position. The optimizer thus produces a set of container-screen pairings for each user, which represent the layout of the merged workspaces. \hl{These pairings are formally defined as $x_{s,c,u}=1$ if screen $s$ is assigned to container $c$ for user $u$, otherwise $x_{s,c,u}=0$}. Our optimization scheme aims to make these two output workspaces as similar as possible.


\subsubsection{Variables}
\textbf{Screens.} \hl{We refer to \emph{``screens''} as the content that may appear on a display, or the display monitor itself. As mentioned, screens can be Shared, Private, or Available. For optimization, we only include screens that belong to the current user, are not physical, and are not assigned during our pre-solving, \emph{or} belong to the remote user, were shared (labeled Shared or Available), and not assigned during pre-solving. We exclude physical screens, as we make the simplifying assumption that they are immovable. This constraint is added for convenience so that users do not have to rearrange their desks when collaborating. Note that if the a physical screen is set to Shared, then its remote equivalent is still passed into our optimizer. Similarly, Available physical screens are sent to the optimizer for potential pre-solving. Finally, we exclude all private screens and treat them as immovable.}


\hl{\textbf{Containers.} Similarly to previous work~\cite{Cheng21SemanticAdapt, Cheng2023InteractionAdapt}, the volumes that indicate usable space are subdivided recursively into a 3D grid of smaller units of containers \hl{of set width $d_c$}, which we refer to as \emph{``voxels''} which represent the containers in the optimization \hl{input (\autoref{tab:input})}. After generation, the following voxels are removed before optimization: 1) voxels that intersect the bounding volume of a physical object; 2) voxels that intersect, occlude, or are occluded by physical or private screens; and 3) voxels that are less than $D_u$ meters away from the user or collaborator. We assign screens to positions by pairing them with a voxel of position $p_c$ representing the screen's bottom-left corner position. The decision variable, $x_{s,c,u}$, only includes valid screen-container-user pairings where the voxel $c$ exists in user $u$'s physical environment and there are enough voxels surrounding $c$ to fit the dimensions ($d_{s}$) of the screen $s$. We keep track of whether each individual voxel is occupied in a bitmap array to help formulate optimization constraints (see Section \ref{sec:constraints}).}

\begin{table}
  \caption{Optimization input parameters.}
  \label{tab:input}
  \begin{tabular}{c p{5cm} l}
    \toprule
    Parameter & Description\\
    \midrule
    $U = \{0, 1\}$ & Set of users in the environment\\
    $z_{avg, u}$ & Average depth distance of screens in user $u$'s workspace\\
    $\mathbf{p}_a \in \mathbb{R}^3$ & Position of avatar relative to $u$\\
    $D_u$ & Threshold distance between user $u$ and their screens\\
    \midrule
    $C_u = \{c_{u1},..., c_{un}\}$ & Set of potential containers for user $u$ to place screens in\\
    $\hl{v_c} \in [0, 1]$ & Semantic utility of placing a screen in container $c$\\
    $\mathbf{p}_c \in \mathbb{R}^3$ & Position of container $c$\\
    $\mathbf{d}_c \in \mathbb{R}^3$ & Dimensions of one voxel $c$\\
    $\mathbf{z}_c \in \mathbb{R}$ & Depth of container $c$ w.r.t user\\
    $c_{occ} = \{c_{u1}, ... \}$ & Set of containers that would be occluded when container $c$ is filled\\
    \midrule
    $S = \{s_1,..., s_n\}$ & Set of screens to integrate into the workspace\\
    $u_s \in \{0, 1\}$ & Which user screen $s$ belongs to\\
    $\hl{v_s} \in [0, 1]$ & Semantic value of screen $s$ in the input environment\\
    $\mathbf{p}_s \in \mathbb{R}^3$ & Position of \hl{bottom-left corner of }screen $s$ in input environment\\
    $\mathbf{d}_s \in \mathbb{R}^3$ & Dimensions of screen $s$\\
    $\mathbf{z}_s \in \mathbb{R}$ & Depth of screen $s$ w.r.t user\\
    \bottomrule
\end{tabular}
\Description{The table has two columns. The first column lists parameters using mathematical notations, and the second column provides their descriptions. Here's the list:
U = {0, 1}: This parameter represents the set of users in the environment.
z_avg, u: This is the average depth distance of screens in user u's workspace.
p_a in R^3: Position of avatar relative to u
D_u: Threshold distance between user u and their screens
The next section details containers for users:
C_u = {c_u1,..., c_un}: Set of potential containers for user u to place screens in
v_c in [0, 1]: Semantic utility of placing a screen in container c
p_c in R^3: Position of container c
d_c in R^3: Dimensions of one voxel c
z_c in R: Depth of container c with respect to the user
c_occ = {c_u1, ... }: Set of containers that would be occluded when container c is filled
The last section describes the screens to be integrated:
S = {s_1,..., s_n}: Set of screens to integrate into the workspace
u_s in {0, 1}: Which user screen s belongs to
v_s in [0, 1]: Semantic value of screen s in the input environment
p_s in R^3: Position of the bottom-left corner of screen s in the input environment
d_s in R^3: Dimensions of screen s
z_s in R: Depth of screen s with respect to the user
The table uses horizontal lines to separate different sections for clarity.}
\end{table}

\subsubsection{Pre-solving}
Before optimization, the system will check on both sides whether there are any physical screens labeled as Available. We assume that physical screens cannot be moved within our optimization model and thus these present a special, simpler case that can be handled without computing our optimization objectives. We justify this by claiming that existing screens are likely to be placed in an already optimal position for the user, and that users tend to prefer physical screens over virtual screens~\cite{Pavanatto21}. Thus, we can improve the efficiency and simplify our computation by first assigning Shared screens to these Available physical screens. We compute the \emph{semantic value} (\hl{$v_s$}, \autoref{eq:semantic_screen}) for these screens as well as for Shared screens. Then, we pair the Shared screen of user A with the Available screen of user B of highest semantic values, respectively, until there are no more physical screens to use or no more screens to share. It is possible that optimization is not necessary after this step.

\begin{figure*}[t!]
    \centering
    \begin{subfigure}[t]{0.24\linewidth}
        \centering
        \includegraphics[height=1.65in]{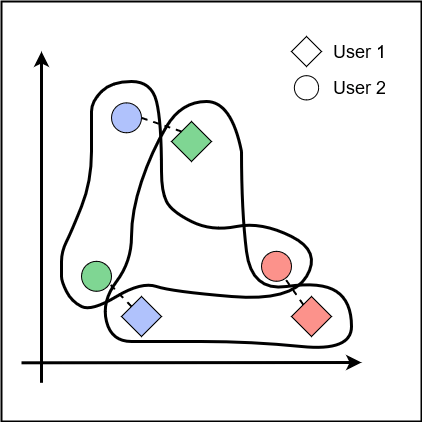}
        \caption{Workspace Agreement}
        \label{fig:workspace_agreement}
    \end{subfigure}
    \hfill
    \begin{subfigure}[t]{0.24\linewidth}
        \centering
        \includegraphics[height=1.65in]{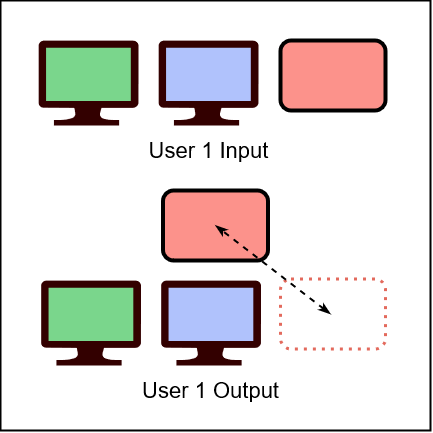}
        \caption{Workspace Modification}
        \label{fig:workspace_modification}
    \end{subfigure}
    \hfill
    \begin{subfigure}[t]{0.24\linewidth}
        \centering
        \includegraphics[height=1.65in]{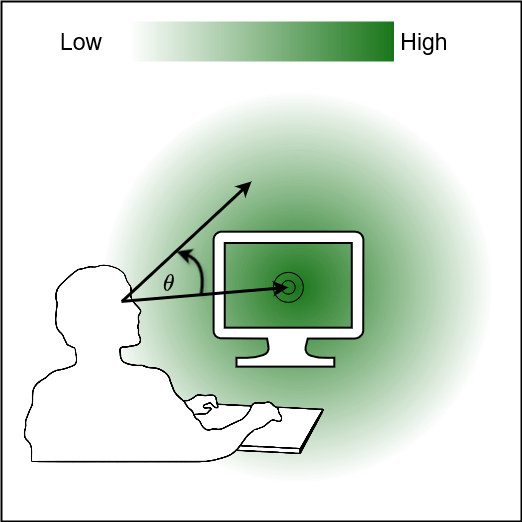}
        \caption{Semantic Utility}
        \label{fig:semantic_utility}
    \end{subfigure}
    \hfill
    \begin{subfigure}[t]{0.24\linewidth}
        \centering
        \includegraphics[height=1.65in]{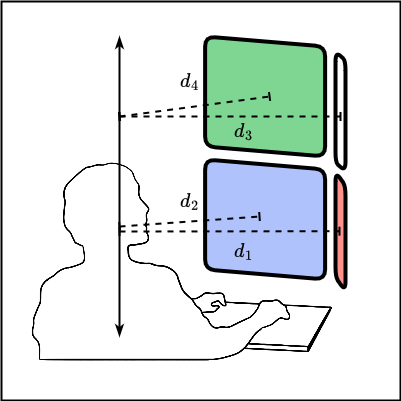}
        \caption{Appearance}
        \label{fig:appearance}
    \end{subfigure}
    \caption{Optimization objectives. \autoref{fig:workspace_agreement} shows two post-integration workspace layouts output for User 1 (diamonds) and User 2 (circles), where markers of the same color represent the same screen. Agreement is calculated as the distance between these clouds. \autoref{fig:workspace_modification} shows an example of workspace modification, which considers only one particular user's workspace prior- and post-optimization. In \autoref{fig:semantic_utility}, positions closer to the workspace center are awarded higher utility, which is calculated using the angular distance $\theta$ between that position and the central direction. Finally, \autoref{fig:appearance} illustrates screen depth, where $z_{avg,u} = average(d_1, d_2, d_3, d_4)$.}
    \Description{a: This image is a diagram with two sets of colored computer monitor icons, labeled "User 1 Input" on the top and "User 1 Output" on the bottom. In the "User 1 Input" section, there are three monitors: one green, one blue, and one red. In the "User 1 Output" section directly below, there is a green and a blue monitor in the same positions as above, but the red monitor has a dashed outline and an arrow pointing from a solid red monitor to its position, indicating a movement.
    b: This image depicts a simple 2D graph with an X-axis and a Y-axis, on which there are two overlapping, irregular closed-loop shapes. Within these shapes, there are six symbols: a blue circle, a green circle a red circle, a red diamond, a green diamon and a blue diamond. Each shape represents a user's workspace. A legend in the upper right corner designates that diamonds represent "User 1" and circles represent "User 2."
    c: This image is a diagram showing a silhouette of a person sitting at a desk facing a computer monitor. Over the monitor is a green overlay with a gradient ranging from low to high, indicated by a bar at the top to show utility.
    d: This illustration presents a schematic view of a person working at a desk with two overlapping virtual screens, one green and one blue, arranged vertically. The screens have dashed lines indicating different depths in space: d1 and d2 for the blue screen, d3 and d4 for the green screen.}
    \label{fig:enter-label}
\end{figure*}
\subsubsection{Objectives}
We define our optimization as the weighted sum of several objectives (\autoref{eq:optimization}) defined as

\hl{
\begin{equation}
    \text{argmin}_{s, c, u} w_a \cdot A + w_m \cdot M + w_v \cdot V + w_p \cdot P.
    \label{eq:optimization}
\end{equation}
}

\textbf{Workspace Agreement (A).} A main goal of our optimization is to minimize inconsistencies between what both users see. Differences in collaborator's awareness of a workspace negatively impacts their ability to communicate~\cite{Gronbaek23}. We refer to this metric as the degree of ``agreement'' of the two workspaces. This agreement is measured by representing each screen as a point using its position vector \hl{$p_s$}, thus interpreting the integrated workspace as a point cloud. The system will output two workspaces, one for User 1 and another for User 2, which are turned into point clouds that contain the common \hl{S}hared screens (see \autoref{fig:workspace_agreement}). The similarity between the two layouts of the integrated workspace is then calculated using Hausdorff distance. Let $d(a, B) = \inf_b d(a, b)$ be the distance of point $a$ to its nearest neighbor in $B$. The Hausdorff distance is defined as the largest such distance:

\begin{equation}
    H(X, Y) = \max\left\{\sup_x d(x, Y), \sup_y d(X, y)\right\}.
\end{equation}

A limitation of this method is that we may lose information about the distance of one screen A in User 1's layout from that \hl{same screen }in User 2's layout, favoring instead an overall shape alignment. This is illustrated in \autoref{fig:workspace_agreement}, since the distance between the neighboring blue-green screen pairs is calculated rather than the green-green or blue-blue pairs, where each color represents a screen. Prioritizing overall shape alignment helps avoid unexpected screen placements when an immovable physical screen is Shared. Future research could experiment with different measures of similarity. Thus, the first objective aims to minimize the following value: 

\begin{equation}
A = H(X_{S, C_{1}, 1}, X_{S, C_2, 2}),
\end{equation}
where $X_{S, C_{1}, 1}$ represents the output set of screen-container pairings for user 1, and $X_{S, C_2, 2}$ represents the output set for user 2.

\textbf{Workspace Modification (M).} Ideally, the workspace of a user should not change drastically when transitioning between individual and collaborative tasks. The initial layout is presumed to be already arranged according to the user's preferences, so making some aspect of the arrangement persistent will preserve some of that inherent value and minimize how much the user has to manually reorganize the workspace after integration~\cite{Niyazov23, Lischke16}. Moreover, this will ensure a smoother transition from the local workspace to the integrated layout, which will reduce task disruptions due to context-switching~\cite{Lindlbauer19}. Similarly to the goal of the Workspace Agreement, the workspace is represented as a point cloud, but this time only considering screens belonging to the current user. Therefore, the two clouds being compared are the workspace prior to and after integration (see \autoref{fig:workspace_modification}) 
\begin{equation}
    M = \frac{1}{2}\left(\sum_{u \in U} H(S_u, X_{S_u, C_u, u})\right),
\end{equation}
where $S_u$ represents the subset of screens $S$ that belong to user $u$, calculated using the ownership input parameter $u_s$ for each screen, and $X_{S_u, C_u, u}$ represents the subset of the output workspace for user $u$ that contains only screens owned by $u$.

\textbf{Semantic Utility \hl{(V)}.} \label{sec:semantic} The placement of screens in a workspace is often done with intention, as each screen serves a particular purpose to the related task. Typically, users will have a central screen that hosts the main focus of the current task and supplementary peripheral screens that are used for quick reference~\cite{Pavanatto21, Lischke16}. Semantic utility refers to how a particular position relates to the purpose of the screen's usage. We approximate semantic utility by giving positions closer to the center of the user's workspace a higher semantic utility score than those further away. This center is determined as the average position of all screens before integration. We assume that this workspace centre is the direction the user is facing when in a resting position. The semantic value of a screen prior to integration is calculated as

\begin{equation}
    \hl{v_s} = e^{-\frac{(\theta-\mu)^2}{2 \sigma^2}}.
    \label{eq:semantic_screen}
\end{equation}
Where $\theta$ is the angular distance between the aforementioned central direction vector from the user's head to the workspace center and the vector pointing from the user's head-mounted display (HMD) to the screen (\autoref{fig:semantic_utility}). The semantic value of a container (\hl{$v_c$}) is calculated similarly to \hl{$v_s$}, but instead considering the vector \hl{from the HMD} to the voxel's position \hl{$p_c$}. Then, when deciding how to place the screen in the integrated workspace, \system{} matches the semantic utility of a particular voxel with the semantic value of the screen. High-utility screens, such as user 1's central screen, should be placed in high-utility containers.  \hl{We negate the final semantic utility value as the optimizer aims to find the minimum
\begin{equation}
    V = -\frac{1}{N_sN_uN_C} \sum_{s} \sum_{u} \sum_{c \in C_u} x_{s, c, u} \left(v_s \cdot v_c\right).
    \label{eq:semantic}
\end{equation}
}

\textbf{Appearance (P).} Our final objective's goal is to ensure that the final workspace has a uniform layout where all the screens are at a similar depth to the user. Constricting screen depth avoids layouts where the screens are positioned too close to the user, which is detrimental to their ability to interact with all screens, as well as being placed too far to be legible. Maintaining a consistent screen depth also avoids uncomfortable visual refocusing when looking between screens and encourages the common layout where screens are placed around the user in a concave shape. Thus, at the start of optimization, we record the average depth distance between the screens and the user's HMD, $z_{avg, u}$ (see \autoref{fig:appearance}), and the depth between each potential voxel and the user's HMD  ($z_c$). Ideally, the chosen voxels would have a depth close to $z_{avg, u}$:
\begin{equation}
    P = \frac{1}{N_sN_uN_C} \sum_{s} \sum_{u} \sum_{c \in C_u} x_{s, c, u} \cdot |z_{avg, u} - z_c|.
    \label{eq:appearance}
\end{equation}

\subsubsection{Constraints}\label{sec:constraints}
In addition to the objectives, we also define a set of constraints that the final result must satisfy. First, every potential container $c$ should contain only one screen $s$ at most in each user $u$'s view. This prevents occlusion between screens, so that no two screens are placed in the same position:
\begin{equation}
    \forall u \in U, \; \forall c \in C_u, \; \sum_{s \in S} x_{s,c,u} \leq 1.
\end{equation}
Second, every shared screen $s$ should only be placed once for each user $u$, leaving no duplicate screens in the scene:
\begin{equation}
    \forall u \in U, \; \forall s \in S, \; \sum_{c \in C_u} x_{s,c,u} = 1.
\end{equation}
Finally, to avoid screens blocking the view of other screens, we use ray-casting to find the set of containers $c_{occ}$ that the current container $c$ would occlude when filled. This container $c$ and the occluded containers $c_{occ}$ should not be occupied simultaneously:
\begin{equation}
    \forall u \in U, \forall c \in C_u, \; \sum_{s} x_{s,c,u} + \sum_{s}\sum_{c_{occ}} x_{s, c_{occ}, u} \leq 1.
\end{equation}

\subsection{Gesture Mapping}

\begin{figure}[t!]
    \centering
    \includegraphics[width=\columnwidth]{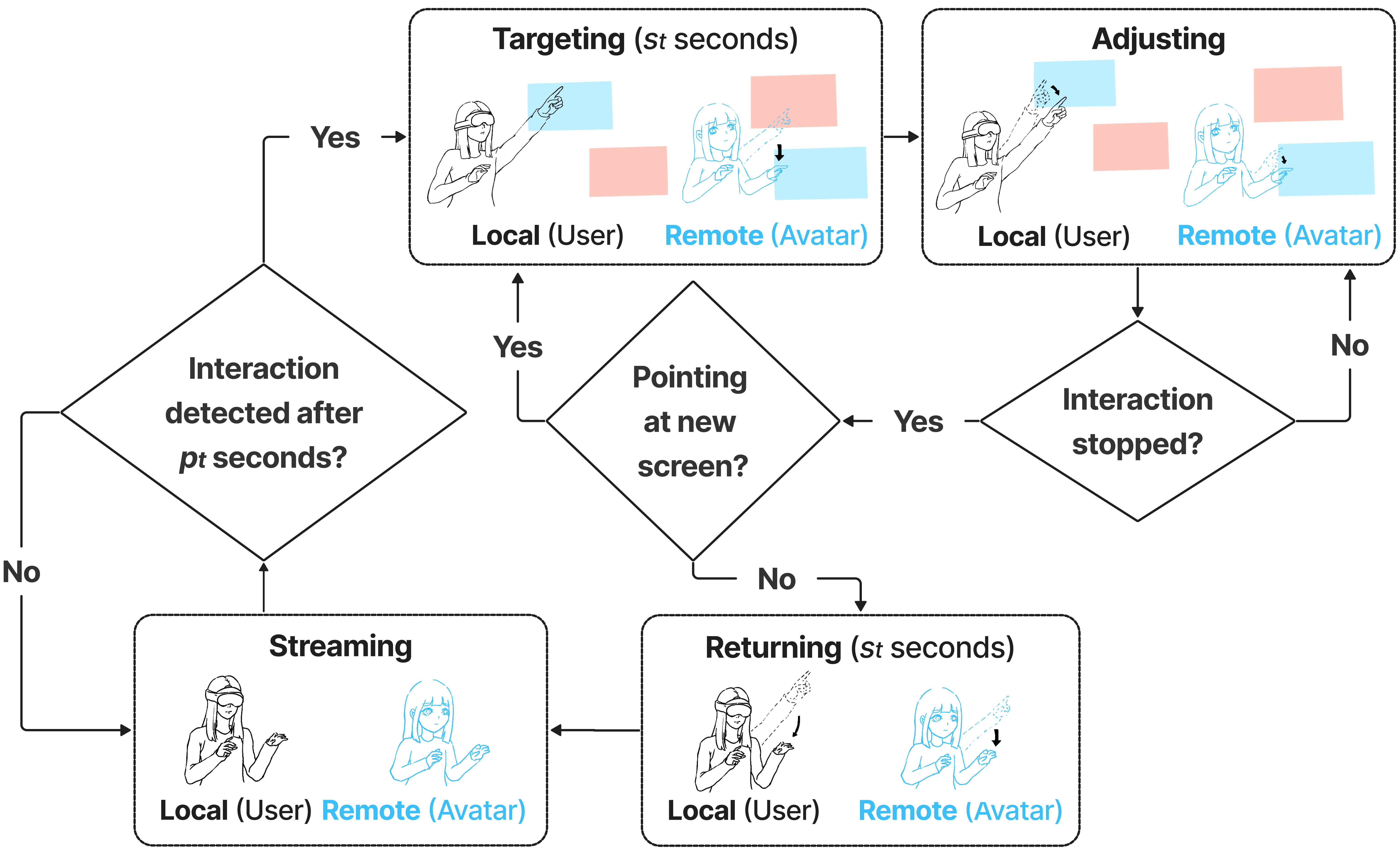}
    \caption{\hl{Retargeting states and conditions.}}
    \label{fig:retarget}
    \Description{The image is a flowchart describing a process involving a local user and a remote avatar interacting in a sequence of steps. It starts with a decision diamond asking if interaction is detected after "pt seconds." If not, the process moves to "Streaming," where illustrations show the local user and the remote avatar side by side, both with VR headsets. If interaction is detected, the process moves to "Targeting (St seconds)" where the local user is pointing at a colored block, and the avatar mimics this action. Next is "Adjusting," where the local user adjusts their pointing, and the avatar follows suit. The flow then leads to another decision diamond: "Interaction changed?" If yes, the process loops back to "Targeting." If no, it advances to another decision diamond: "Interaction stopped?" If no, the flow goes back to "Adjusting." If yes, the process moves to "Returning (St seconds)" where the local user and avatar are shown with a dotted line retreating from the interaction, completing the cycle. Each step has time elements involved, and the process iterates based on the user's continuous interaction or the stoppage thereof. The diagram is used to explain a VR interaction retargeting mechanism.}
\end{figure}

After finding the optimal screen placement, the workspace of each user is adjusted. Furthermore, the optimization scheme provides information on the corresponding screens on each user's side along with the position of the remote collaborator relative to the local user (i.e., right or left side). Since it is unlikely that the final result will have two perfectly aligned workspaces due to physical constraints, we need to augment the remote user's avatar to ensure that any pointing interaction and gaze remain consistent between workspaces and meaningful for collaboration.

The main issue arises when the screens have different layouts after optimization. As a result, the remote user may appear to be pointing into thin air. Previous works have developed inverse kinematics systems that translate any pointing and gaze interaction to its intended target~\cite{Yoon21, Kang2023Retargeting}. We extend these works by considering when multiple independent screens are present in the workspace. This introduces a new challenge when the user performs a gesture pointing from one screen to another. If the relative position of the two screens is not considered, such a gesture may result in ``snapping'' or discontinuous motion. The same applies to user avatars: from User A's view, User B is on their right, where as from User B's view, User A is on their right. If the remote avatar matches the movements of the user while User B is looking at User A, User B's avatar may appear to be looking away from User A's point of view.

Our retargeting algorithm transitions between multiple states to ensure stable avatar movements (\autoref{fig:retarget}). If the user is not interacting with a common screen or the collaborator's remote avatar, their remote avatar is in a \emph{Streaming} state where the user's position is directly applied to the avatar without any retargeting. \hl{During streaming, the remote avatars' relative positions to the local users are considered. If both remote avatars are on the same side relative to the local user (e.g., right side), the streaming positions are mirrored to ensure spatial consistency. When an interaction with a screen is detected (i.e., gazed on or pointed at), the avatar transitions to a \emph{Targeting} state where the avatar position is interpolated over $s_t$ seconds to point at the corresponding screen position that the user is pointing to. When the avatar has reached the interacted object, the avatar enters an \emph{Adjusting} state where the avatar is adjusted to point to the corresponding point on the screen to which the user points. If the user stops an interaction, the avatar transitions to a \emph{Returning} state where the avatar position is interpolated back to the normal streaming position over $s_t$ seconds. If the user changes the interaction by pointing at a new screen, the avatar will instead transition into the Targeting state towards the new screen.}

\hl{Interaction is detected when the user's hand or gaze is pointing toward one of the shared screens or the remote user's avatar for longer than a specified threshold $p_t$~\cite{Yoon21}. Although introducing a delay at the start of interaction, this threshold ensures that transitions between states remain stable, which is important in cases where screens are densely placed or when tracking is poor. For gaze pointing, we use the head direction of the user. For hand pointing, we create a ray from the user's head position that passes through the hand position. After an interaction is detected, a spatial hysteresis is applied to enlarge the boundary of the interaction area by a scale factor $b_{scale}$ to cover cases where the user is pointing or looking at the object edge~\cite{Hansen2018GazeInteraction, Zhao2007Earpod}. Any screen that is interacted with is also highlighted to the collaborator by displaying a red border around the screen to help guide attention to the interacted screen. During all retargeting states we calculate the positions and rotations necessary to point at the corresponding position. For the hands, we create a ray from the head towards the interaction position and place the hand $D_r$ meters along the ray to minimize the chance that the avatar intersects with a screen during retargeted pointing. After the retargeted positions and rotations are calculated, we apply inverse kinematics to the avatar to create a coherent avatar posture.}

\subsection{Implementation}\label{seq:implementation}
We implemented the system in Unity. We used the Meta Quest Pro as the HMD for both users. We used the HMD's pass-through to view the physical environment. Due to the limited resolution of the pass-through image, virtual screens were overlayed on top of physical screens to ensure visibility of screen content. With better pass-through resolution and ability to accurately display bright screens we envisage this step to not be necessary. We used the Meta hand-tracking to track user positions for retargeting. Furthermore, we used Meta avatars and their inverse kinematics for retargeting. Finally, we used the ``uDesktopDuplication'' package to include screen content in the MR environment. We used the ``Mirror'' Networking package for all communication between users via UDP. Finally, we implemented the optimization model using Gurobi through a Python interface. Communication between Unity and Gurobi was done via a websocket.
\section{Study}
\begin{figure*}
    \centering
    \includegraphics[width=1\textwidth]{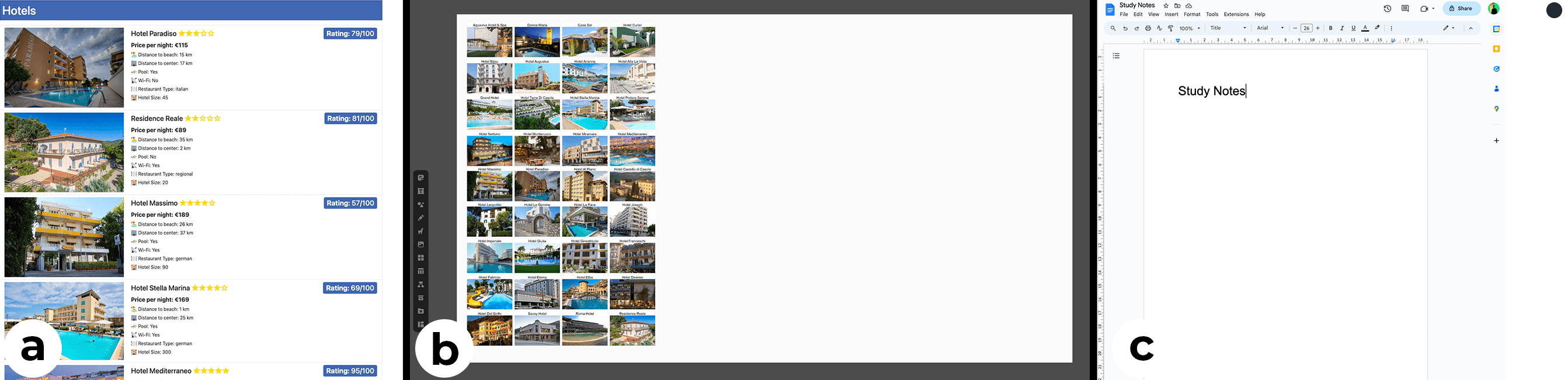}
    \caption{Study apps. a) List webpage of hotels. b) Mural whiteboard of hotels for sense-making. c) Google doc for note-taking.}
    \Description{This image displays a three-part digital study setup, labeled (a), (b), and (c), representing different applications for organizing study materials related to hotels. a) Shows a list webpage of hotels with detailed information such as name, price per night, ratings, distance to beach, city center, and amenities like Wi-Fi and restaurant availability. Each listing has a photo, a description, and a highlighted user rating. b) Depicts a Mural whiteboard or similar visual organizing application with a grid of hotel images that seem to be used for categorizing and sense-making, allowing the user to visually organize information about different hotels. c) Is a Google Docs window titled "Study Notes," which is a blank document with the cursor at the beginning, ready for the user to take notes.}
    \label{fig:study-apps}
\end{figure*}

\begin{figure}[t]
    \centering
    \includegraphics[width=1\linewidth]{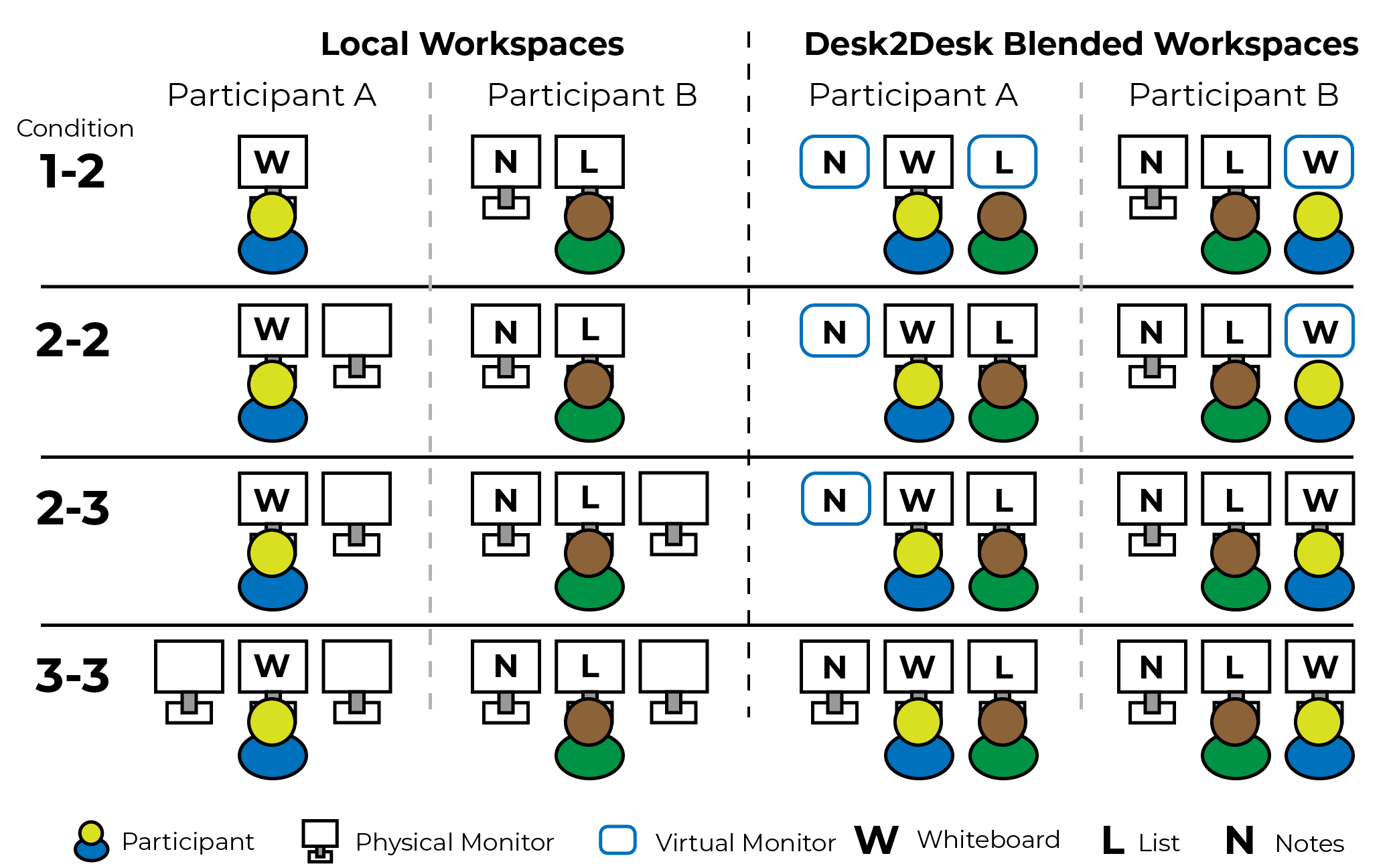}
    \caption{Workspace combinations for user study.}
    \Description{This image is a grid layout showing workspace combinations for a user study with two participants, labeled A and B. It illustrates various conditions (1-2, 2-2, 2-3, 3-3) in two distinct workspace settings: individual "Workspaces" and "Blended Workspaces". Each row represents a different condition and shows the arrangement of tools used by each participant. The tools are symbolized by icons: a "W" for Whiteboard, an "L" for List, an "N" for Notes, and monitors represented by squares - a physical one outlined and a virtual one filled in blue. In the "Workspaces" section, Participant A consistently has a Whiteboard (W) and a virtual monitor, while Participant B has a List (L), Notes (N), and a physical monitor. In the "Blended Workspaces" section, the tools are rearranged between participants for collaboration, with the virtual and physical monitors being exchanged and the Whiteboard, List, and Notes being shared or swapped. The icon key at the bottom identifies a blue circle as the participant, a physical monitor as an outlined square, and a virtual monitor as a filled blue square.}
    \label{fig:study-layouts}
\end{figure}

We conducted a user study to investigate \system{}'s suitability for remote side-by-side collaboration during a typical desktop task. The aim of the study was to study user behavior while using our system, and to understand how \system{} can facilitate deictic gestures and speech and its impact on the quality of collaboration. This was done in the context of a ``planning and negotiation'' task where users had to book a hotel. We compared two integration modes: \system{}, and a ``Clone'' baseline as a within-subject independent variable. Furthermore, we varied the combination of workspaces as a between-subject independent variable (\autoref{fig:study-layouts}). 

\subsection{Task}
The choice of study task was based on the criteria of 1) requiring discussion and negotiation for task completion to encourage collaboration; 2) be desktop-based; 3) requiring the use of multiple monitors; 4) can be adapted to fit the study setting (e.g., control of task duration); 5) be related to real-world tasks; and 6) do not require specialized prior knowledge. As such, we decided on a hotel search task from previous work on collaboration on tabletops~\cite{Jetter2011FaceStreams} and AR~\cite{Fink2022ReLocations}. In the hotel search task, the participants must decide on accommodation for a mutual vacation. Each participant is given a different set of accommodation criteria (e.g., ``The hotel should have a pool'' or ``The hotel should have at least 4 stars''). However, the task is designed so that no hotel meets all criteria. Therefore, participants must discuss, negotiate, and consider their criteria to find the ``optimal'' solution. The tasks represent a controllable representative of collaborative activities, where the collaborators must compare options, negotiate, and agree on a final solution. Furthermore, awareness of each other's activity and attention is beneficial, but not required, to solve the task.

We adopted the task in the following ways. Two participants worked together in separate rooms and workspaces. The workspaces differed in the number of monitors, and their layout (cf. \autoref{sec:study-environment}). The collaborators had at their disposal a custom website (\autoref{fig:study-apps}a) containing a list of hotels and their information (e.g., price, number of stars), a shared online Mural\footnote{\url{https://www.mural.co/}} whiteboard with attached images of hotels for sense-making (\autoref{fig:study-apps}b), and a blank shared Google Doc\footnote{\url{https://docs.google.com/}} to take notes (\autoref{fig:study-apps}c). To solve the task, participants had to agree on a hotel from a set of 36 fictive hotels. We adopted the same set of hotels and criteria as in previous work by \citet{Fink2022ReLocations}. At the beginning of the task, participants received a printed sheet with their personal criteria (four for each participant). As no hotel met all criteria, participants had to negotiate, discuss, and agree on a final hotel. Participants were asked to make as few concessions as possible. Participants were informed that they could organize the whiteboard, take notes, and use the hotel list as they wished. Participants were asked to keep a single window open on each monitor. The participants notified the examiners when they had agreed on a hotel. We kept a time limit of 15 minutes to keep the whole study within a reasonable time and to ensure consistency between pairs and integration modes.

\subsection{Study Environments and Apparatus} \label{sec:study-environment}

\begin{figure}[t]
    \centering
    \includegraphics[width=1\columnwidth]{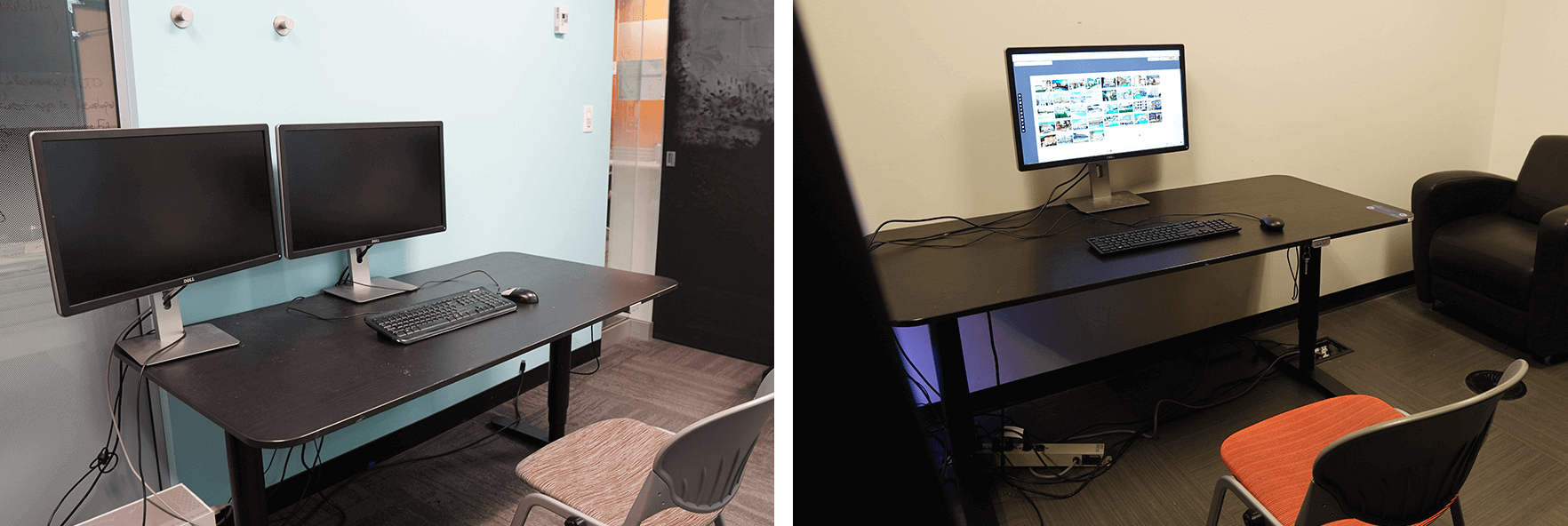}
    \caption{The two physical environments used for the study. The figure illustrates the 1-2 condition.}
    \Description{The image shows two different physical environments used for a study, illustrating the '1-2 condition'. On the left, the setup includes a black desk with two turned-off Dell monitors placed side by side, a keyboard, a mouse, and a grey chair. The wall behind the desk features a whiteboard with some writing on it, and there's a glimpse of a glass panel partially covered with notes. On the right, the setup is a single black desk with one turned-on monitor displaying multiple images, possibly a gallery or a planning tool. There's also a keyboard, a mouse, an orange chair with a black back, and a black sofa in the background. The room has a neutral color scheme and appears to be a small, private office or study room.}
    \label{fig:environment}
\end{figure}
We used two rooms for the study to represent the remote setting. Each room was equipped with a workspace. Each workspace had an office desk and chair, a desktop computer, a mouse and keyboard, and a Meta Quest Pro (\autoref{fig:environment}). To include a varying set of workspaces to evaluate our system, the physical display configuration varied for each participant. We used four unique combinations of the number of physical monitors between workspaces (\autoref{fig:study-layouts} left). \hl{These layouts were decided by the number of applications required for the task (in total 3 physical screens before integration), and represent the most common layouts found in a previous survey of workspace device configurations~\cite{yuan2022understanding}.} All workspaces had horizontal monitor layouts for the sake of simplicity of study. The workspaces were already defined via the Oculus Room Setup before participants arrived. \hl{In the study, the collaborator's virtual avatar was always placed on the right side of the participant for both participants to  guarantee retargeting during the study}. Finally, the applications were distributed between participants to enable interaction for both encourage collaboration. The list and the note application were always paired to allow easy transfer of information from the list to the note document, and the other participant always had the whiteboard. For \system{}, optimization was performed online when participants connected and merged their workspaces. 

\subsection{Study Integration Modes}

\begin{figure}
    \centering
    \includegraphics[width=1\columnwidth]{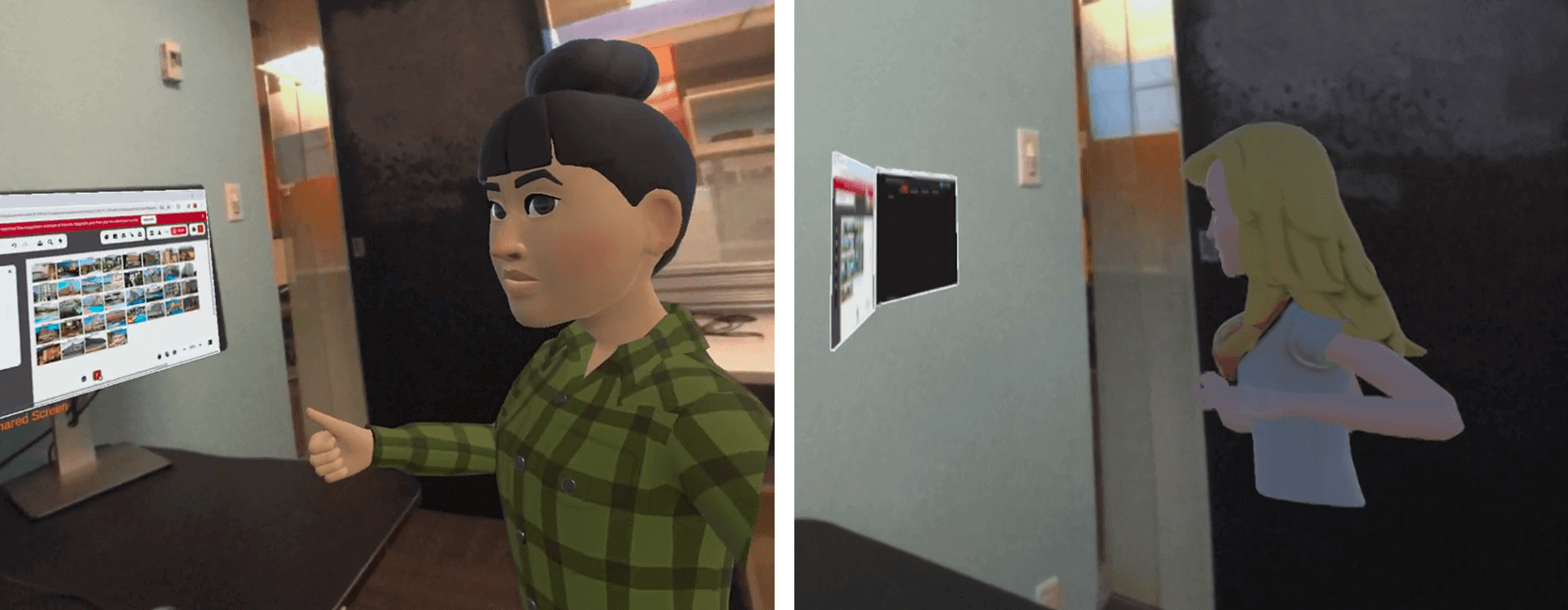}
    \caption{Study integration modes for our study. Left: \system{}. Right: Clone baseline.}
    \Description{This image shows two side-by-side photographs demonstrating different integration modes for a study, with the left labeled "Desk2Desk" and the right labeled "Clone baseline". On the left, "Desk2Desk" features a three-dimensional avatar of a person with a bun, dressed in a green plaid shirt, standing and pointing at a computer screen displaying a webpage with multiple images, representing a shared workspace in a virtual environment. On the right, "Clone baseline" depicts a different three-dimensional avatar with long blonde hair, standing and facing a wall where a small, floating screen showing a similar webpage is displayed, representing a separate virtual workspace.}
    \label{fig:study-conditions}
\end{figure}
We compared two workspace integrations, \system{} and a ``Clone'' baseline (\autoref{fig:study-conditions}).  In the Clone integration baseline, the remote collaborator's whole workspace was spawned next to the local user's workspace to mimic sitting at desks next to each other. In addition to the apps shown in \autoref{fig:study-apps}, participants also had access to the Zoom videoconferencing app that could be used to share screens between users. We opted for an MR-based baseline compared to a typical desktop teleconferencing system as we anticipated that the weight and limited display and passthrough resolution of modern HMDs would have a significant impact on results. We used the following \system{} parameters for the study based on extensive pilot testing: Weightings: $w_a = 0.2$, $w_m = 0.2$, $\hl{w_v} = 0.5$, and $w_p = 0.1$. Voxel parameters: $D_u=0.5m$, $\hl{d_c}=0.1m$. Retargeting parameters: $b_{scale}=1.3$, $s_t=0.5s$, $D_r=0.42m$, $p_t=0.1s$ for hand pointing and $p_t=0.3s$ for head pointing.

\subsection{Procedure and Participants}
We recruited 8 pairs of participants (N=16, 8 male, 8 female) from a local university. The average age of the participants was 23.1 years (SD = 1.5). Participants were welcomed and brought to a common room. After being informed about the study, participants signed a consent form, completed a demographic questionnaire and chose one of 36 standard Meta Avatars as their virtual representation. The study was approved by the university ethics board. All participants rated their individual experience with AR/VR HMDs (e.g., Meta Quest or Microsoft HoloLens) and teleconferencing systems (e.g., Zoom or Skype) from ``never'' to ``daily'', as well as their familiarity with their collaborator (from ``met first time'' to ``very familiar'') on a 5-point Likert scale. Among the 16 participants, 9 had tried AR/VR HMDs ``on occasion'', while 7 had ``never tried it before''; 11 participants used teleconferencing systems on a ``weekly'' basis, 3 on a ``daily'' basis and 2 ``on occasion''; 8 participants met their collaborator for the ``first time'', 5 were ``very familiar'', 2 were ``acquainted with'', and 1 was ``familiar'' with their study partner. 

The participants were then separately introduced to their designated workspace, the study task, and the task applications. The participant then put on the HMD, and the examiner introduced the first integration condition. The workspaces were then merged by participants pressing a virtual ``Join'' button when ready, the examiners ensured that the collaborators could hear and see each other, and the participants received their task descriptions with individual hotel criteria on a sheet of paper. The participants then completed the task with one of the integration conditions. After 15 minutes of performing the task, participants answered a post-task questionnaire containing the Raw NASA Task load index~\cite{Sandra2006Nasa}, System Usability Scale (SUS)~\cite{Bangor2008sus}, and spatial and social components of the Temple Presence Inventory (TPI)~\cite{Lombard2009tpi} to gather perceived workload, usability, and presence, respectively, along with a set of questions regarding their collaboration and workspace layout. The participants then completed the task again with the second integration condition and a second set of personal hotel criteria. The integration condition was counterbalanced with a balanced Latin square. After completing the task with both integration modes, the examiners conducted semi-structured interviews about their experiences and opinions. The interviews were conducted separately to ensure that they were unaware of any differences in workspace layout to minimize bias. The study took on average 90 minutes to complete and participants were compensated 25 CAD.

\subsection{Analysis}

\begin{table*}[t]
    \centering
    \caption{Communication coding during task based on previous work on attention cue visualizations~\cite{Dangelo17GazeProgramming, Bovo2023Cov}.}
    \resizebox{\linewidth}{!}{
        \begin{tabular}{rl}
             \textbf{(a) Context} &  Reference to a monitor based on its location within the workspace. For example ``Top Left Corner''.\\
             \textbf{(b) Select}& Implicit verbal reference by pointing with the cursor or highlighting text or elements\\
             \textbf{(b) Pointer}& Implicit verbal reference by pointing with their hand to highlight a monitor or element directly\\
             & and optionally utter words like ``this'', ``over here.'', or directly mentioning ``where I’m pointing''.\\
             (\textbf{c) User Relative}& Reference to an element in relation to the frame of reference of the other user, such as: ``on your right/left side'', \\
             &``close to/far from you'', ``above/below you''.\\
             \textbf{(d) Deictic}& Deictic reference such as ``this'', ``that ''or ``here'', or when a participant directly refers to their gaze, for example,\\ 
             &``Right where I am looking''.\\
             \textbf{(e) Temporal}& Reference to an element previously referenced, such as ``before, after, or earlier''.\\
             \textbf{(f) Element}& Explicit reference to an element on the monitor.\\
             \textbf{(g) Sequential}& Reference to a monitor or element on the monitor by naming a number representing the element, for example, \\
             &``Shall we move to the third monitor''. An alternative expression might consist of the participant suggesting \\
             &moving the collaboration focus to the next or previous monitor/element.\\
             \textbf{(h) Color}& Reference to an element by naming its color.\\
        \end{tabular}
    }
    \label{tab:behavior-coding}
    \Description{The table outlines various methods of communication coding used during tasks, referencing previous studies on attention cue visualizations.
(a) Context Reference: Refers to a monitor based on its location within the workspace, such as "Top Left Corner".
(b) Select: Involves implicit verbal references made by pointing with the cursor or highlighting text or elements on the screen.
(b) Pointer: Includes implicit verbal references made by pointing with their hand to a monitor or element directly, and optionally saying words like "this", "over here", or "where I’m pointing".
(c) User Relative: Pertains to references made in relation to the other user’s frame of reference, for example, "on your right/left side", "close to/far from you", or "above/below you".
(d) Deictic: Involves using deixis in communication, such as "this", "that", or "here", or when a participant directly refers to where they are looking, like "Right where I am looking".
(e) Temporal: References an element that was mentioned earlier in the conversation with terms like "before", "after", or "earlier".
(f) Element: Makes an explicit reference to a specific element visible on the monitor.
(g) Sequential: Refers to a monitor or an element on it by naming a sequence number, such as "Shall we move to the third monitor", or by suggesting moving the focus to the next or previous item.
(h) Color: Identifies an element by naming its color.}
\end{table*}
As the goal of \system{} is to encourage side-by-side collaboration, we analyzed participants behavior during collaboration, answers from semi-structured interviews and questionnaires. To understand how workspace integration influenced communication between pairs, all video and audio recordings and transcripts were coded for references and acknowledgments. Each time a participant made an acknowledged reference, the type and time were recorded in the transcript. Reference codes (\autoref{tab:behavior-coding}) were based on previous work that aimed to improve remote pairwise collaboration using visual attention cues~\cite{Bovo2023Cov, Dangelo17GazeProgramming}. Two coders performed the analysis. Each transcribed trial was analyzed by one coder and then reviewed by the other. The roles between coders rotated for each trial. After coding, we counted the number of occurrences that each pair of participants made under each experimental condition.
\section{Results}

The participants were able to complete the task within the allotted time limit for both \system{} and Clone. As such, we structure our results based on user behavior during collaboration, and subjective feedback from questionnaire and interview results. 
\subsection{Speech and Gesture Results}

\begin{figure*}
    \centering
    \includegraphics[width=\linewidth]{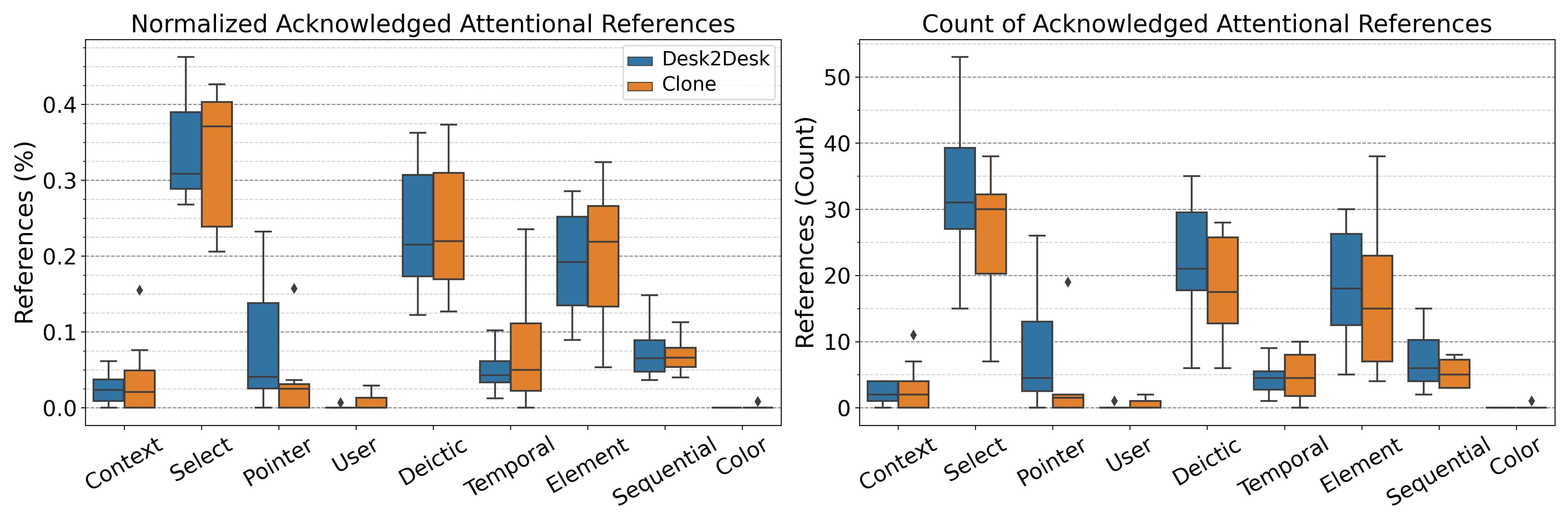}
    \caption{Fraction (left) and count (right) of each reference type uttered by the pair in the two experimental conditions.}
    \label{fig:references}
    \Description{The image presents two side-by-side boxplot graphs showing data from a study comparing two experimental conditions: "Desk2Desk" and "Clone". On the left, the graph titled "Normalized Acknowledged Attentional References" displays the fraction (in percentage) of different reference types (Context, Select, Pointer, User, Deictic, Temporal, Element, Sequential, Color) uttered by the pair of participants. Each reference type has two boxplots, one for each condition (Desk2Desk in blue, Clone in orange), illustrating the spread and median of the data. On the right, the graph titled "Count of Acknowledged Attentional References" shows the raw count of each reference type for the same conditions and categories as the left graph. The vertical axes indicate the number of references, and each category's boxplot displays the range, median, and outliers.}
\end{figure*}

As participant pairs worked together on finding the hotel, each had to make references to different hotels, notes, screens and imaged across the workspaces. We counted the number of utterances of each reference type from \autoref{tab:behavior-coding} as they completed the tasks. As each pair performed the same set of tasks within the same timeframe (15 minutes) we also normalized the values to produce a fraction of each reference type. The count and distribution of references can be seen in \autoref{fig:references}. Wilcoxon signed ranked tests between study conditions showed that participants performed more pointing gestures in \system{} \hl{($z$=$2.12$, $p$=$.034$, $r$=$.61$)}, highlighting how integrating the workspaces combined with side-by-side collaboration encourages users to physically point to indicate attention. Although participants tended to make more references in \system{}, we found no other statistically significant differences. 

\subsection{Questionnaire Results}
\begin{figure}
    \centering
    \includegraphics[width=1\columnwidth]{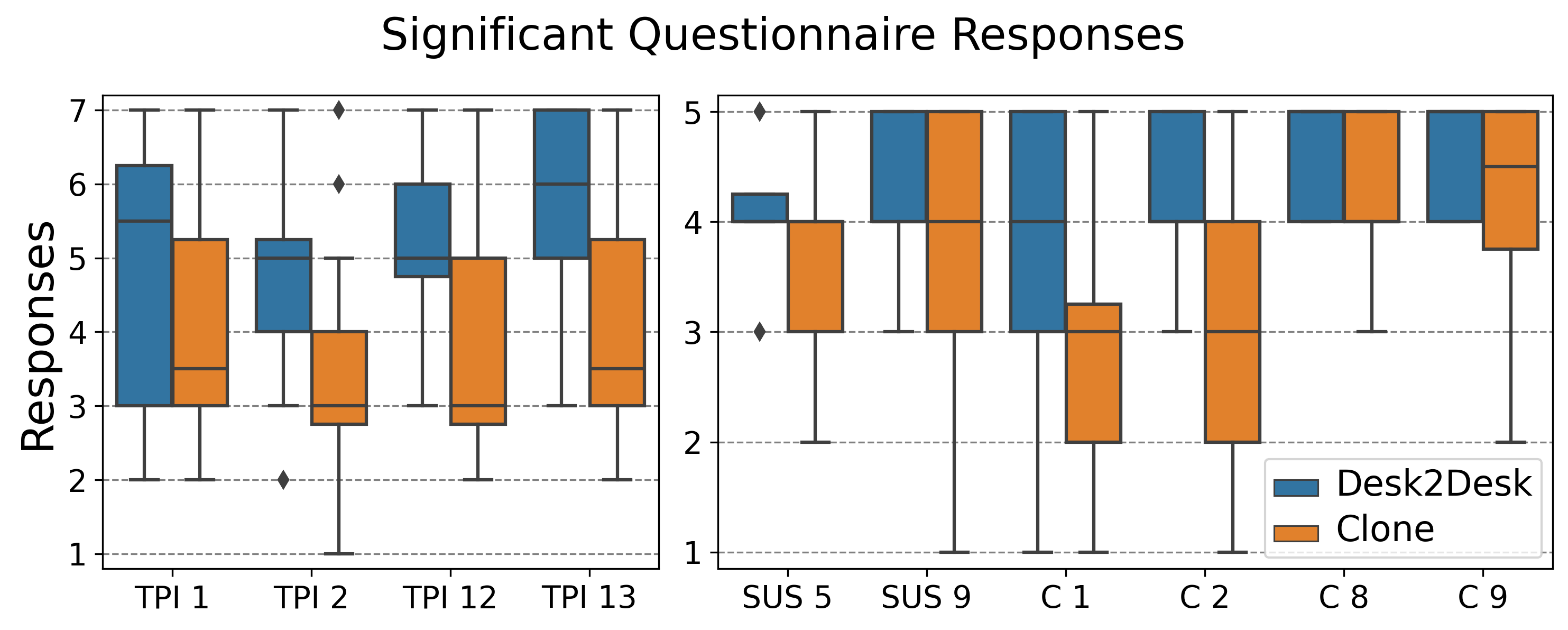}
    \caption{\hl{Quartiles of significant questionnaire responses.}}
    \label{fig:questionnaire}
    \Description{The image is a bar chart titled "Significant Questionnaire Responses", displaying responses to selected questionnaire items (Q1, Q2, Q12, Q13, Q19, Q23, Q31, Q32, Q38, Q39) for two different conditions: Desk2Desk (in blue) and Clone (in orange). Each question is represented on the horizontal axis and is associated with two bars side by side, one for each condition. The vertical axis measures the response level on a scale presumably from 1 to 7, as the highest bar reaches just below 7. The bars indicate the average response to each question, with error bars that likely represent the variability or standard deviation of the responses.
}
\end{figure}

We collected responses on TPI, NASA TLX, SUS and workspace, and collaboration-based questionnaires. In this section, we include the questions that were found significant through Wilcoxon signed ranked tests (\autoref{fig:questionnaire}). For all questionnaire results, please refer to the supplemental material.

Questionnaire results were generally in favour of \system{} over Clone. For spatial presence questions, participants felt significantly more that ``objects and people you saw/heard had come to the place you were'' (\hl{TPI 1: $z$=$2.41$, $p$=$.016$, $r$=$.69$}) in \system{} ($M$=$5.0$, $SD$=$1.75$) compared to Clone ($M$=$4.19$, $SD$=$1.68$). Tests also showed a significant difference (\hl{TPI 2: $z$=$2.79$, $p$=$.005$, $r$=$.81$}) in favour of \system{} ($M$=$4.81$, $SD$=$1.56$) over Clone ($M$=$3.50$, $SD$=$1.55$) when it comes to their feeling of ``they could reach out and touch the objects or people they saw/heard''.

For social presence TPI questions, Wilcoxon tests showed significance (\hl{TPI 12: $z$=$2.82$, $p$=$.004$, $r$=$.81$}) in favor of \system{} ($M$=$5.13$, $SD$=$1.09$) over Clone ($M$=$3.87$, $SD$=$1.75$) when asked about their level of ``control over the interaction with the person or people they saw/heard''. Furthermore participants felt that it felt significantly more ``like someone they saw/heard in the environment was talking directly to them'' for \system{} (\hl{TPI 13: $z$=$2.98$, $p$=$.003$, $r$=$.86, M$=$5.75$, $SD$=$1.24$}) over Clone ($M$=$4.19$, $SD$=$1.83$).

SUS results showed significance (\hl{SUS 5: $z$=$2.65$, $p$=$.008$, $r$=$.77$}) in favor of \system{} ($M$=$4.06$, $SD$=$.68$) over Clone ($M$=$3.62$, $SD$=$.72$) for how well ``functions in the system are integrated''. Results also showed they felt more ``confident'' using \system{} (\hl{SUS 9: $z$=$2.23$, $p$=$.026$, $r$=$.64, M$=$4.06$, $SD$=$.68$}) over Clone ($M$=$3.62$, $SD$=$.719$).

For questions regarding their collaboration, participant felt more like ``they were working side-by-side'' for \system{} (\hl{C 1: $z$=$2.63$, $p$=$.009$, $r$=$.76, M$=$3.75$, $SD$=$1.18$}) over Clone ($M$=$2.88$, $SD$=$1.03$). Participants also reported that they felt that they were ``performing the tasks in the same environment'' for \system{} (\hl{C 2: $z$=$2.92$, $p$=$.004$, $r$=$.84, M$=$4.19$, $SD$=$.75$}) over Clone ($M$=$2.94$, $SD$=$1.12$). When asked whether they felt that ``my partner could understand my actions'' they indicated that \system{} was significantly better (\hl{C 8: $z$=$.2.45$, $p$=$.014$, $r$=$.71, M$=$4.68$, $SD$=$.48$}) over Clone ($M$=$4.31$, $SD$=$.70$). Finally, participants said it was significantly easier to ``understand what was happening in the workspace'' for \system{} (\hl{C 9: $z$=$2.12$, $p$=$.034$, $r$=$.64, M$=$4.69$, $SD$=$.48$}) over Clone ($M$=$4.12$, $SD$=$1.09$).

\subsection{Interview Results}
At the end of the study we performed a semi-structured interview to ask them about their opinions about the two integration modes, the collaborator visualizations, and the workspace layout.

\subsubsection{Comparing Integration Modes}

When asked to compare the two integration modes, a majority (15 of 16) of participants preferred \system{} over the Clone baseline. Participants mentioned that \system{} made them feel more collaborative: \textit{``it gives me a sense that we are working together''} (P3\_2) and \textit{``it feels like someone sitting next to you like a real person''} (P2\_2). Participants also highlighted the capability of dynamically adding and placing screens: \textit{``I really like \system{} because you can add more screens. [..] It's just way more usable. It feels like it really increases our productivity''} (P8\_2) and \textit{``I can see both my screen and the other person's screen. That's much better than if you are just doing screen share''} (P7\_2). Finally participants mentioned that it was easier to focus on the same part of the workspace with \system{}. P4\_2 mentioned \textit{``since we are focussing on the same screen, it's easier for us to focus on the same task and understand what each other mean''}. Participants who performed the task with a lower amount of screens mentioned the usefullness of being able to add additional virtual displays in comparison to sharing over Zoom. P8\_1 stated that \textit{``When there are multiple screens it is much easier to know if my partner is looking at the same screen. This is much easier for us to communicate with each other to see what we are trying to talk about''}. P5\_1 confirmed this view by stating that \textit{``Being able to see my partner's screen at the same time I can see my screen is a very good benefit''}. One participant mentioned that they prefer the baseline due to their familiarity with Zoom. P7\_1 stated `\textit{`It was similar to working with Zoom, I did not feel like adding more screens was that much efficient''}. 

\subsubsection{Collaborator Visualizations}

When asked about their collaborator's avatar and the screen highlighting the responses were varied. A subset of participants enjoyed their collaborator's avatar. P6\_2 stated \textit{``I liked the partner's avatar which is right next to me. It actually made me feel like we were  working together''} and P5\_1 added \textit{``I think the avatar is great. I can see not only the avatar's face but also the hand movement''}. Finally, (P4\_2) stated \textit{``When I'm wondering what my partner is doing I look at the avatar''}. However, participants also mentioned that they did not pay attention to the avatar due to the side-by-side collaborative nature. P7\_2 mentioned that \textit{``I think I only focus on the screens''} while P6\_1 added \textit{``The person sitting next to me does not make me complete the task faster, but it is probably giving me some emotional support''}. Participants also mentioned that the desktop setup reduced the need for physical interaction between participants\textit{``Because we are also using the mouse, sometimes there is no need to physically point at the screen''} (P6\_2). Finally, participants mentioned the cartoon style of the avatar, \textit{``It was too cartoonish to feel immersive. With eye tracking and more realism it would feel like I'm working with a real person''} (P2\_2).

Regarding the screen highlighting, participants mentioned that it was beneficial for collaboration and guiding attention. P2\_2 mentioned that \textit{``If my partner wants to talk about a screen, I can see the highlight''}. Participants especially thought the highlighting was beneficial together with mouse movements as the highlighting represents the general area of attention while the mouse is used for further refinement. P7\_2 mention that \textit{``The screen highlighting only shows the current screen. [..] For more detail we would look at the mouse''}. P6\_1 added \textit{``I felt aware of my partner because I could see the mouse moving and whether she is looking at my screen or not''}.

\subsubsection{Workspace Layout}
\hl{
When asked about the screen layouts generated by \system{}, participants mentioned that the screens were \textit{``very clearly laid out''} (P3\_1) and that \textit{``The layout is kind of typical, like it's very comfortable''} (P6\_1). Participants further mentioned that the layout was comfortable for them  \textit{``I like having vertical and horizontal screens. [..] It feels very comfortable to use.''} (P8\_2), \textit{``The screens are compact and you can see all screens at one.''} (P7\_2), and ``the screen setting is familiar so I don't need to think why the layout looks like this.'' by P6\_1. 
}
\section{Discussion}

In this paper, we contributed \system{}, an optimization-based approach to merge two MR workspaces to enable remote side-by-side collaboration that accounts for each user's local physical environment. \system{} enables the integration of significantly varying physical environments, as virtual screens can be adaptively spawned and moved to fill the ``gaps'' between workspaces. Any resulting inconsistencies caused by physical limitations and differences between user environments are handled by retargeting to ensure consistency in gaze and pointing directions. Fundamentally, our approach enables both collaborators the sensation of working side-by-side with someone from their own private workspace.

Our study results showed that participants could efficiently collaborate between varying workspace combinations compared to a baseline based on Zoom. This highlights the utility of our optimization approach, as participants mentioned that the generated workspaces felt comfortable and natural. A further benefit of the MR workspace is that it can effectively utilize vertical free space which are underutilized by physical screens. Participants mentioned that utilizing space above the physical screen was good, as it reduced the need for back-and-forth head movements caused by wide horizontal workspaces. Furthermore, our pre-solving approach makes use of existing, unutilized physical screen to keep workspaces decluttered. 

A fundamental benefit of \system{} is its ability to adaptively spawn additional screens to enable multi-screen sharing without occluding users' own screens. This feature was especially highlighted by participants with only a single screen, who mentioned that they became more passive during the baseline, as they had to choose between looking at their own screen content or their collaborator's shared screen. \system{} also enables gaze and pointing gestures as additional cues for collaboration. Although they may not be necessary for side-by-side collaboration, participants mentioned that it increased immersion and their feeling of collaboration. 

\begin{figure}
    \centering
    \includegraphics[width=\columnwidth]{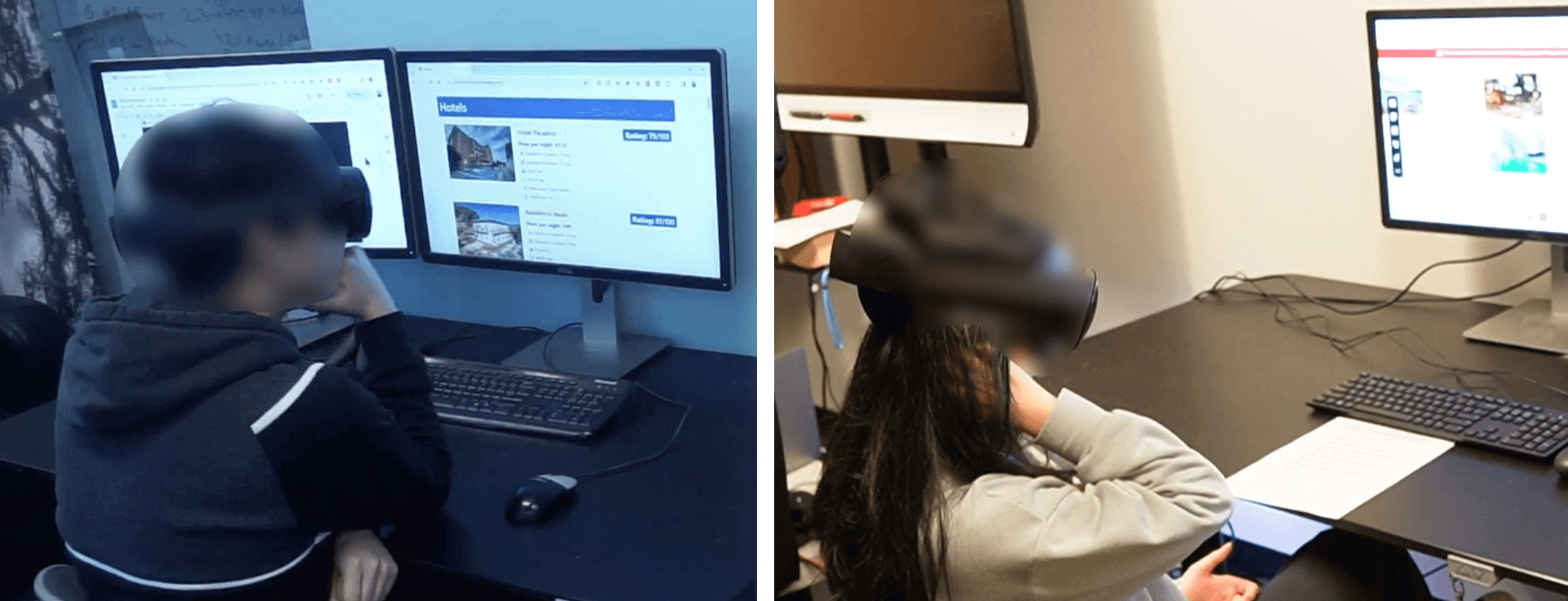}
    \caption{Participant hand postures during the study were often outside of the tracking range.}
    \label{fig:postures}
    \Description{The image shows two people at desks, each wearing a VR headset and looking at computer screens. Left Side: A person in a black and gray hoodie is seated at a desk with two monitors displaying hotel listings. They rest their hand on their cheek while using the VR headset. Right Side: A person in a gray sweater with long dark hair sits at a desk with a monitor displaying a webpage. They also rest their hand on their cheek while wearing a VR headset, with a keyboard, mouse, and a paper in front of them.}
\end{figure}

Our retargeting approach is essential in dealing with the unavoidable physical misalignment between workspaces. The participants mentioned that the avatars helped them get immersed in the collaboration and the frequent use of pointing gestures during study tasks showcased its utility. Furthermore, multiple participants mentioned that they referred to the collaborator's head direction to understand their current attention and actions. However, it became apparent from the study that hand-tracking struggle in workspace settings. As seen in \autoref{fig:postures}, participants would frequently move their hands to a comfortable position beyond the hand-tracking area (e.g., behind their head, sitting on or leaning on them). This would cause the avatar arms to move into unexpected positions. However, this was not noticed by participants, as focus was mainly directed towards the displays.

\hl{The balance between maximizing workspace consistency between users while minimizing workspace modification and maximizing utility is an important trade-off exposed by our objectives. Through pilot testing, we found that prioritizing individual workspaces was more beneficial even with increased retargeting. However, with tasks that require more User Relative references (e.g. ``on your left''), it is reasonable to speculate that minimizing retargeting should be prioritized. Furthermore, our work prioritized overall shape-alignment of workspaces rather than individual screen-pairings between users. We believe that prioritizing shape-alignment avoids unexpected screen placements in cases of sharing immovable physical screens. However, User Relative and Context (e.g. ``Top left corner'') references then become less useful and may cause confusion. Future work should investigate these trade-offs and supporting more communication cues to further collaboration.}

Finally, our optimization approach of prioritizing the semantic utility of screens, the consistency of screen positions between users, and minimizing changes to the local MR workspace is complementary to other work that has investigated adaptive interfaces using, for example, cognitive load~\cite{Lindlbauer19}, social acceptability~\cite{Medeiros2022PassengerWorkspace}, and ergonomics~\cite{Belo2021XRgonomics}. A key contrast to these works is that our adaptation is done for multiple users. This opens up a significant design space where collaborative and multi-user objectives and constraints can be designed and evaluated to create desirable and novel interface behaviors in multi-user settings.

\subsection{Future Work and Limitations}
Our work can be expanded in multiple directions. Although our system allows for the sharing of screens, only the sharer is able to interact with the screen. Multiple participants mentioned that more integration would be beneficial, i.e., allowing both users to interact with shared screens. This would also increase the feeling of working together side by side. Another avenue of expansion is to incorporate common movable physical devices such as laptops and tablets into the optimization. Currently, we assume that the physical objects are not changeable and that the workspace remains stationary after integration. Being able to continuously update the workspace layout would enable more dynamic collaborative behaviors and more flexibility in supported workspace layouts. Furthermore, our system is limited to typical desktop windows operated by mouse and keyboard. Extending to merge these windows with virtual windows that can be interactable via VR interaction like direct touch could be a future avenue for expansion. Finally, our system could be expanded beyond two users to support workspace-based collaborative tasks with additional users (e.g. mob programming).

The main technical limitation of our work is the limited resolution of the HMD and pass-through. Most modern pass-through HMDs struggle to show physical display content. Therefore, we overlaid virtual windows on top of the physical screens to minimize the effect of this limitation. Even then, we needed to significantly lower the display resolution (1280$\times$720) than is typically supported by physical screens. This was a major participant complaint mentioned during our user study. As pass-through and display technology improves, we expect this to become less of an issue.

Finally, we believe that our approach would benefit from additional controlled studies and longitudinal evaluations. \hl{There are an almost infinite number of layout combinations, of which we only formally evaluated a few. Future studies can focus on integrating the vast permutations of possible workspace layouts~\cite{yuan2022understanding}, and extreme use cases (e.g., when there is not enough space for all screens).} Furthermore, it would be interesting to longitudinally study our system as a replacement for typical videoconferencing systems such as Zoom. Previous research has shown that working completely in VR can be demanding~\cite{Biener2022VrWeek}. However, one can imagine that the user only wears our headset when doing certain collaborative remote meetings. We believe that such studies would inspire future adaptive and collaborative systems.
\section{Conclusion} 
We have presented \system{}, an optimization-based approach for integrating remote MR workspaces for side-by-side collaboration. Our novel approach shows that established optimization techniques for layout optimization can be extended to collaborative scenarios. Results from our user study show that adaptive workspaces are effective for providing collaborators with shared information and to support collaboration and immersion, especially in workspaces with few physical screens. We believe that our work opens up new avenues for collaborative adaptive systems that integrate virtual workspaces and interfaces with familiar physical counterparts.

\bibliographystyle{ACM-Reference-Format}
\bibliography{bibliography}


\begin{thebibliography}{74}


\ifx \showCODEN    \undefined \def \showCODEN     #1{\unskip}     \fi
\ifx \showDOI      \undefined \def \showDOI       #1{#1}\fi
\ifx \showISBNx    \undefined \def \showISBNx     #1{\unskip}     \fi
\ifx \showISBNxiii \undefined \def \showISBNxiii  #1{\unskip}     \fi
\ifx \showISSN     \undefined \def \showISSN      #1{\unskip}     \fi
\ifx \showLCCN     \undefined \def \showLCCN      #1{\unskip}     \fi
\ifx \shownote     \undefined \def \shownote      #1{#1}          \fi
\ifx \showarticletitle \undefined \def \showarticletitle #1{#1}   \fi
\ifx \showURL      \undefined \def \showURL       {\relax}        \fi
\providecommand\bibfield[2]{#2}
\providecommand\bibinfo[2]{#2}
\providecommand\natexlab[1]{#1}
\providecommand\showeprint[2][]{arXiv:#2}

\bibitem[Bai et~al\mbox{.}(2020)]%
        {Bai2020Collaboration}
\bibfield{author}{\bibinfo{person}{Huidong Bai}, \bibinfo{person}{Prasanth
  Sasikumar}, \bibinfo{person}{Jing Yang}, {and} \bibinfo{person}{Mark
  Billinghurst}.} \bibinfo{year}{2020}\natexlab{}.
\newblock \showarticletitle{A User Study on Mixed Reality Remote Collaboration
  with Eye Gaze and Hand Gesture Sharing}. In
  \bibinfo{booktitle}{\emph{Proceedings of the 2020 CHI Conference on Human
  Factors in Computing Systems}} (Honolulu, HI, USA)
  \emph{(\bibinfo{series}{CHI '20})}. \bibinfo{publisher}{Association for
  Computing Machinery}, \bibinfo{address}{New York, NY, USA},
  \bibinfo{pages}{1–13}.
\newblock
\showISBNx{9781450367080}
\urldef\tempurl%
\url{https://doi.org/10.1145/3313831.3376550}
\showDOI{\tempurl}


\bibitem[Bangor et~al\mbox{.}(2008)]%
        {Bangor2008sus}
\bibfield{author}{\bibinfo{person}{Aaron Bangor}, \bibinfo{person}{Philip~T.
  Kortum}, {and} \bibinfo{person}{Miller~James T.}}
  \bibinfo{year}{2008}\natexlab{}.
\newblock \showarticletitle{An Empirical Evaluation of the System Usability
  Scale}.
\newblock \bibinfo{journal}{\emph{International Journal of Human–Computer
  Interaction}} \bibinfo{volume}{24}, \bibinfo{number}{6}
  (\bibinfo{year}{2008}), \bibinfo{pages}{574--594}.
\newblock
\urldef\tempurl%
\url{https://doi.org/10.1080/10447310802205776}
\showDOI{\tempurl}


\bibitem[Biener et~al\mbox{.}(2022)]%
        {Biener2022VrWeek}
\bibfield{author}{\bibinfo{person}{Verena Biener}, \bibinfo{person}{Snehanjali
  Kalamkar}, \bibinfo{person}{Negar Nouri}, \bibinfo{person}{Eyal Ofek},
  \bibinfo{person}{Michel Pahud}, \bibinfo{person}{John~J. Dudley},
  \bibinfo{person}{Jinghui Hu}, \bibinfo{person}{Per~Ola Kristensson},
  \bibinfo{person}{Maheshya Weerasinghe}, \bibinfo{person}{Klen~Čopič
  Pucihar}, \bibinfo{person}{Matjaž Kljun}, \bibinfo{person}{Stephan
  Streuber}, {and} \bibinfo{person}{Jens Grubert}.}
  \bibinfo{year}{2022}\natexlab{}.
\newblock \showarticletitle{Quantifying the Effects of Working in VR for One
  Week}.
\newblock \bibinfo{journal}{\emph{IEEE Transactions on Visualization and
  Computer Graphics}} \bibinfo{volume}{28}, \bibinfo{number}{11}
  (\bibinfo{year}{2022}), \bibinfo{pages}{3810--3820}.
\newblock
\urldef\tempurl%
\url{https://doi.org/10.1109/TVCG.2022.3203103}
\showDOI{\tempurl}


\bibitem[Biener et~al\mbox{.}(2020)]%
        {Biener2020BreakingTheScreen}
\bibfield{author}{\bibinfo{person}{Verena Biener}, \bibinfo{person}{Daniel
  Schneider}, \bibinfo{person}{Travis Gesslein}, \bibinfo{person}{Alexander
  Otte}, \bibinfo{person}{Bastian Kuth}, \bibinfo{person}{Per~Ola Kristensson},
  \bibinfo{person}{Eyal Ofek}, \bibinfo{person}{Michel Pahud}, {and}
  \bibinfo{person}{Jens Grubert}.} \bibinfo{year}{2020}\natexlab{}.
\newblock \showarticletitle{Breaking the Screen: Interaction Across Touchscreen
  Boundaries in Virtual Reality for Mobile Knowledge Workers}.
\newblock \bibinfo{journal}{\emph{IEEE Transactions on Visualization \&
  Computer Graphics}} \bibinfo{volume}{26}, \bibinfo{number}{12}
  (\bibinfo{date}{dec} \bibinfo{year}{2020}), \bibinfo{pages}{3490--3502}.
\newblock
\showISSN{1941-0506}
\urldef\tempurl%
\url{https://doi.org/10.1109/TVCG.2020.3023567}
\showDOI{\tempurl}


\bibitem[Bovo et~al\mbox{.}(2023)]%
        {Bovo2023Cov}
\bibfield{author}{\bibinfo{person}{Riccardo Bovo}, \bibinfo{person}{Daniele
  Giunchi}, \bibinfo{person}{Ludwig Sidenmark}, \bibinfo{person}{Joshua Newn},
  \bibinfo{person}{Hans Gellersen}, \bibinfo{person}{Enrico Costanza}, {and}
  \bibinfo{person}{Thomas Heinis}.} \bibinfo{year}{2023}\natexlab{}.
\newblock \showarticletitle{Speech-Augmented Cone-of-Vision for Exploratory
  Data Analysis}. In \bibinfo{booktitle}{\emph{Proceedings of the 2023 CHI
  Conference on Human Factors in Computing Systems}} (Hamburg, Germany)
  \emph{(\bibinfo{series}{CHI '23})}. \bibinfo{publisher}{Association for
  Computing Machinery}, \bibinfo{address}{New York, NY, USA}, Article
  \bibinfo{articleno}{162}, \bibinfo{numpages}{18}~pages.
\newblock
\showISBNx{9781450394215}
\urldef\tempurl%
\url{https://doi.org/10.1145/3544548.3581283}
\showDOI{\tempurl}


\bibitem[Burgoon and Bacue(2003)]%
        {burgoon2003nonverbal}
\bibfield{author}{\bibinfo{person}{Judee~K Burgoon} {and}
  \bibinfo{person}{Aaron~E Bacue}.} \bibinfo{year}{2003}\natexlab{}.
\newblock \showarticletitle{Nonverbal communication skills}.
\newblock \bibinfo{journal}{\emph{Handbook of communication and social
  interaction skills}} (\bibinfo{year}{2003}), \bibinfo{pages}{179--219}.
\newblock


\bibitem[Cheng et~al\mbox{.}(2021)]%
        {Cheng21SemanticAdapt}
\bibfield{author}{\bibinfo{person}{Yifei Cheng}, \bibinfo{person}{Yukang Yan},
  \bibinfo{person}{Xin Yi}, \bibinfo{person}{Yuanchun Shi}, {and}
  \bibinfo{person}{David Lindlbauer}.} \bibinfo{year}{2021}\natexlab{}.
\newblock \showarticletitle{SemanticAdapt: Optimization-Based Adaptation of
  Mixed Reality Layouts Leveraging Virtual-Physical Semantic Connections}. In
  \bibinfo{booktitle}{\emph{The 34th Annual ACM Symposium on User Interface
  Software and Technology}} (Virtual Event, USA) \emph{(\bibinfo{series}{UIST
  '21})}. \bibinfo{publisher}{Association for Computing Machinery},
  \bibinfo{address}{New York, NY, USA}, \bibinfo{pages}{282–297}.
\newblock
\showISBNx{9781450386357}
\urldef\tempurl%
\url{https://doi.org/10.1145/3472749.3474750}
\showDOI{\tempurl}


\bibitem[Cheng et~al\mbox{.}(2023)]%
        {Cheng2023InteractionAdapt}
\bibfield{author}{\bibinfo{person}{Yi~Fei Cheng}, \bibinfo{person}{Christoph
  Gebhardt}, {and} \bibinfo{person}{Christian Holz}.}
  \bibinfo{year}{2023}\natexlab{}.
\newblock \showarticletitle{InteractionAdapt: Interaction-Driven Workspace
  Adaptation for Situated Virtual Reality Environments}. In
  \bibinfo{booktitle}{\emph{Proceedings of the 36th Annual ACM Symposium on
  User Interface Software and Technology}} (San Francisco, CA, USA)
  \emph{(\bibinfo{series}{UIST '23})}. \bibinfo{publisher}{Association for
  Computing Machinery}, \bibinfo{address}{New York, NY, USA}, Article
  \bibinfo{articleno}{109}, \bibinfo{numpages}{14}~pages.
\newblock
\showISBNx{9798400701320}
\urldef\tempurl%
\url{https://doi.org/10.1145/3586183.3606717}
\showDOI{\tempurl}


\bibitem[Clark and Brennan(1991)]%
        {clark1991grounding}
\bibfield{author}{\bibinfo{person}{Herbert~H. Clark} {and}
  \bibinfo{person}{Susan~E. Brennan}.} \bibinfo{year}{1991}\natexlab{}.
\newblock \showarticletitle{Grounding in communication.}
\newblock \bibinfo{journal}{\emph{Perspectives on socially shared cognition}}
  (\bibinfo{year}{1991}), \bibinfo{pages}{127--149}.
\newblock
\showISBNx{1-55798-121-3 (Hardcover)}
\urldef\tempurl%
\url{https://doi.org/10.1037/10096-006}
\showDOI{\tempurl}


\bibitem[Collins(2011)]%
        {Collins2011Accounting}
\bibfield{author}{\bibinfo{person}{J~Carlton Collins}.}
  \bibinfo{year}{2011}\natexlab{}.
\newblock \showarticletitle{Increase productivity with multiple monitors}.
\newblock \bibinfo{journal}{\emph{J. Account}} (\bibinfo{year}{2011}).
\newblock


\bibitem[Czerwinski et~al\mbox{.}(2003)]%
        {Czerwinski2003Workspace}
\bibfield{author}{\bibinfo{person}{Mary Czerwinski}, \bibinfo{person}{Greg
  Smith}, \bibinfo{person}{Tim Regan}, \bibinfo{person}{Brian Meyers},
  \bibinfo{person}{George~G. Robertson}, {and} \bibinfo{person}{Gary~K.
  Starkweather}.} \bibinfo{year}{2003}\natexlab{}.
\newblock \showarticletitle{Toward Characterizing the Productivity Benefits of
  Very Large Displays}. In \bibinfo{booktitle}{\emph{Human-Computer Interaction
  {INTERACT} '03: {IFIP} {TC13} International Conference on Human-Computer
  Interaction, 1st-5th September 2003, Zurich, Switzerland}},
  \bibfield{editor}{\bibinfo{person}{Matthias Rauterberg},
  \bibinfo{person}{Marino Menozzi}, {and} \bibinfo{person}{Janet Wesson}}
  (Eds.). \bibinfo{publisher}{{IOS} Press}.
\newblock


\bibitem[D'Angelo and Begel(2017)]%
        {Dangelo17GazeProgramming}
\bibfield{author}{\bibinfo{person}{Sarah D'Angelo} {and}
  \bibinfo{person}{Andrew Begel}.} \bibinfo{year}{2017}\natexlab{}.
\newblock \showarticletitle{Improving Communication Between Pair Programmers
  Using Shared Gaze Awareness}. In \bibinfo{booktitle}{\emph{Proceedings of the
  2017 CHI Conference on Human Factors in Computing Systems}} (Denver,
  Colorado, USA) \emph{(\bibinfo{series}{CHI '17})}.
  \bibinfo{publisher}{Association for Computing Machinery},
  \bibinfo{address}{New York, NY, USA}, \bibinfo{pages}{6245–6290}.
\newblock
\showISBNx{9781450346559}
\urldef\tempurl%
\url{https://doi.org/10.1145/3025453.3025573}
\showDOI{\tempurl}


\bibitem[Efran(1968)]%
        {Efran1968Collaboration}
\bibfield{author}{\bibinfo{person}{Jay~S. Efran}.}
  \bibinfo{year}{1968}\natexlab{}.
\newblock \showarticletitle{Looking for approval: Effects on visual behavior of
  approbation from persons differing in importance.}
\newblock \bibinfo{journal}{\emph{Journal of Personality and Social
  Psychology}} \bibinfo{volume}{10}, \bibinfo{number}{1}
  (\bibinfo{year}{1968}), \bibinfo{pages}{21--25}.
\newblock
\urldef\tempurl%
\url{https://doi.org/10.1037/h0026383}
\showDOI{\tempurl}


\bibitem[Endert et~al\mbox{.}(2012)]%
        {Endert2012Workspace}
\bibfield{author}{\bibinfo{person}{Alex Endert}, \bibinfo{person}{Lauren
  Bradel}, \bibinfo{person}{Jessica Zeitz}, \bibinfo{person}{Christopher
  Andrews}, {and} \bibinfo{person}{Chris North}.}
  \bibinfo{year}{2012}\natexlab{}.
\newblock \showarticletitle{Designing Large High-Resolution Display
  Workspaces}. In \bibinfo{booktitle}{\emph{Proceedings of the International
  Working Conference on Advanced Visual Interfaces}} (Capri Island, Italy)
  \emph{(\bibinfo{series}{AVI '12})}. \bibinfo{publisher}{Association for
  Computing Machinery}, \bibinfo{address}{New York, NY, USA},
  \bibinfo{pages}{58–65}.
\newblock
\showISBNx{9781450312875}
\urldef\tempurl%
\url{https://doi.org/10.1145/2254556.2254570}
\showDOI{\tempurl}


\bibitem[Evangelista~Belo et~al\mbox{.}(2021)]%
        {Belo2021XRgonomics}
\bibfield{author}{\bibinfo{person}{Jo\~{a}o~Marcelo Evangelista~Belo},
  \bibinfo{person}{Anna~Maria Feit}, \bibinfo{person}{Tiare Feuchtner}, {and}
  \bibinfo{person}{Kaj Gr\o{}nb\ae{}k}.} \bibinfo{year}{2021}\natexlab{}.
\newblock \showarticletitle{XRgonomics: Facilitating the Creation of Ergonomic
  3D Interfaces}. In \bibinfo{booktitle}{\emph{Proceedings of the 2021 CHI
  Conference on Human Factors in Computing Systems}} (, Yokohama, Japan,)
  \emph{(\bibinfo{series}{CHI '21})}. \bibinfo{publisher}{Association for
  Computing Machinery}, \bibinfo{address}{New York, NY, USA}, Article
  \bibinfo{articleno}{290}, \bibinfo{numpages}{11}~pages.
\newblock
\showISBNx{9781450380966}
\urldef\tempurl%
\url{https://doi.org/10.1145/3411764.3445349}
\showDOI{\tempurl}


\bibitem[Evangelista~Belo et~al\mbox{.}(2022)]%
        {Belo2022Auit}
\bibfield{author}{\bibinfo{person}{Jo\~{a}o~Marcelo Evangelista~Belo},
  \bibinfo{person}{Mathias~N. Lystb\ae{}k}, \bibinfo{person}{Anna~Maria Feit},
  \bibinfo{person}{Ken Pfeuffer}, \bibinfo{person}{Peter K\'{a}n},
  \bibinfo{person}{Antti Oulasvirta}, {and} \bibinfo{person}{Kaj
  Gr\o{}nb\ae{}k}.} \bibinfo{year}{2022}\natexlab{}.
\newblock \showarticletitle{AUIT – the Adaptive User Interfaces Toolkit for
  Designing XR Applications}. In \bibinfo{booktitle}{\emph{Proceedings of the
  35th Annual ACM Symposium on User Interface Software and Technology}} (Bend,
  OR, USA) \emph{(\bibinfo{series}{UIST '22})}. \bibinfo{publisher}{Association
  for Computing Machinery}, \bibinfo{address}{New York, NY, USA}, Article
  \bibinfo{articleno}{48}, \bibinfo{numpages}{16}~pages.
\newblock
\showISBNx{9781450393201}
\urldef\tempurl%
\url{https://doi.org/10.1145/3526113.3545651}
\showDOI{\tempurl}


\bibitem[Fender et~al\mbox{.}(2017)]%
        {Fender2017Heatspace}
\bibfield{author}{\bibinfo{person}{Andreas Fender}, \bibinfo{person}{David
  Lindlbauer}, \bibinfo{person}{Philipp Herholz}, \bibinfo{person}{Marc Alexa},
  {and} \bibinfo{person}{J\"{o}rg M\"{u}ller}.}
  \bibinfo{year}{2017}\natexlab{}.
\newblock \showarticletitle{HeatSpace: Automatic Placement of Displays by
  Empirical Analysis of User Behavior}. In
  \bibinfo{booktitle}{\emph{Proceedings of the 30th Annual ACM Symposium on
  User Interface Software and Technology}} (Qu\'{e}bec City, QC, Canada)
  \emph{(\bibinfo{series}{UIST '17})}. \bibinfo{publisher}{Association for
  Computing Machinery}, \bibinfo{address}{New York, NY, USA},
  \bibinfo{pages}{611–621}.
\newblock
\showISBNx{9781450349819}
\urldef\tempurl%
\url{https://doi.org/10.1145/3126594.3126621}
\showDOI{\tempurl}


\bibitem[Fidalgo et~al\mbox{.}(2023)]%
        {Fidalgo23}
\bibfield{author}{\bibinfo{person}{Catarina~G. Fidalgo},
  \bibinfo{person}{Maurício Sousa}, \bibinfo{person}{Daniel Mendes},
  \bibinfo{person}{Rafael~Kuffner Dos~Anjos}, \bibinfo{person}{Daniel
  Medeiros}, \bibinfo{person}{Karan Singh}, {and} \bibinfo{person}{Joaquim
  Jorge}.} \bibinfo{year}{2023}\natexlab{}.
\newblock \showarticletitle{MAGIC: Manipulating Avatars and Gestures to Improve
  Remote Collaboration}. In \bibinfo{booktitle}{\emph{2023 IEEE Conference
  Virtual Reality and 3D User Interfaces (VR)}}. \bibinfo{pages}{438--448}.
\newblock
\urldef\tempurl%
\url{https://doi.org/10.1109/VR55154.2023.00059}
\showDOI{\tempurl}


\bibitem[Fink et~al\mbox{.}(2022)]%
        {Fink2022ReLocations}
\bibfield{author}{\bibinfo{person}{Daniel~Immanuel Fink},
  \bibinfo{person}{Johannes Zagermann}, \bibinfo{person}{Harald Reiterer},
  {and} \bibinfo{person}{Hans-Christian Jetter}.}
  \bibinfo{year}{2022}\natexlab{}.
\newblock \showarticletitle{Re-Locations: Augmenting Personal and Shared
  Workspaces to Support Remote Collaboration in Incongruent Spaces}.
\newblock \bibinfo{journal}{\emph{Proc. ACM Hum.-Comput. Interact.}}
  \bibinfo{volume}{6}, \bibinfo{number}{ISS}, Article \bibinfo{articleno}{556}
  (\bibinfo{date}{nov} \bibinfo{year}{2022}), \bibinfo{numpages}{30}~pages.
\newblock
\urldef\tempurl%
\url{https://doi.org/10.1145/3567709}
\showDOI{\tempurl}


\bibitem[Fröhler et~al\mbox{.}(2022)]%
        {Frohler2022Xva}
\bibfield{author}{\bibinfo{person}{B. Fröhler}, \bibinfo{person}{C. Anthes},
  \bibinfo{person}{F. Pointecker}, \bibinfo{person}{J. Friedl},
  \bibinfo{person}{D. Schwajda}, \bibinfo{person}{A. Riegler},
  \bibinfo{person}{S. Tripathi}, \bibinfo{person}{C. Holzmann},
  \bibinfo{person}{M. Brunner}, \bibinfo{person}{H. Jodlbauer},
  \bibinfo{person}{H.-C. Jetter}, {and} \bibinfo{person}{C. Heinzl}.}
  \bibinfo{year}{2022}\natexlab{}.
\newblock \showarticletitle{A Survey on Cross-Virtuality Analytics}.
\newblock \bibinfo{journal}{\emph{Computer Graphics Forum}}
  \bibinfo{volume}{41}, \bibinfo{number}{1} (\bibinfo{year}{2022}),
  \bibinfo{pages}{465--494}.
\newblock
\urldef\tempurl%
\url{https://doi.org/10.1111/cgf.14447}
\showDOI{\tempurl}


\bibitem[Gr\o{}nb\ae{}k et~al\mbox{.}(2023)]%
        {Gronbaek23}
\bibfield{author}{\bibinfo{person}{Jens Emil~Sloth Gr\o{}nb\ae{}k},
  \bibinfo{person}{Ken Pfeuffer}, \bibinfo{person}{Eduardo Velloso},
  \bibinfo{person}{Morten Astrup}, \bibinfo{person}{Melanie
  Isabel~S\o{}nderk\ae{}r Pedersen}, \bibinfo{person}{Martin Kj\ae{}r},
  \bibinfo{person}{Germ\'{a}n Leiva}, {and} \bibinfo{person}{Hans Gellersen}.}
  \bibinfo{year}{2023}\natexlab{}.
\newblock \showarticletitle{Partially Blended Realities: Aligning Dissimilar
  Spaces for Distributed Mixed Reality Meetings}. In
  \bibinfo{booktitle}{\emph{Proceedings of the 2023 CHI Conference on Human
  Factors in Computing Systems}} (Hamburg, Germany) \emph{(\bibinfo{series}{CHI
  '23})}. \bibinfo{publisher}{Association for Computing Machinery},
  \bibinfo{address}{New York, NY, USA}, Article \bibinfo{articleno}{456},
  \bibinfo{numpages}{16}~pages.
\newblock
\urldef\tempurl%
\url{https://doi.org/10.1145/3544548.3581515}
\showDOI{\tempurl}


\bibitem[Gr\o{}nb\ae{}k et~al\mbox{.}(2024)]%
        {Groenbaek2024Whiteboard}
\bibfield{author}{\bibinfo{person}{Jens Emil~Sloth Gr\o{}nb\ae{}k},
  \bibinfo{person}{Juan S\'{a}nchez~Esquivel}, \bibinfo{person}{Germ\'{a}n
  Leiva}, \bibinfo{person}{Eduardo Velloso}, \bibinfo{person}{Hans Gellersen},
  {and} \bibinfo{person}{Ken Pfeuffer}.} \bibinfo{year}{2024}\natexlab{}.
\newblock \showarticletitle{Blended Whiteboard: Physicality and
  Reconfigurability in Remote Mixed Reality Collaboration}. In
  \bibinfo{booktitle}{\emph{Proceedings of the CHI Conference on Human Factors
  in Computing Systems}} (Honolulu, HI, USA) \emph{(\bibinfo{series}{CHI
  '24})}. \bibinfo{publisher}{Association for Computing Machinery},
  \bibinfo{address}{New York, NY, USA}, Article \bibinfo{articleno}{798},
  \bibinfo{numpages}{16}~pages.
\newblock
\showISBNx{9798400703300}
\urldef\tempurl%
\url{https://doi.org/10.1145/3613904.3642293}
\showDOI{\tempurl}


\bibitem[Grudin(2001)]%
        {Grudin2001MultiDisplay}
\bibfield{author}{\bibinfo{person}{Jonathan Grudin}.}
  \bibinfo{year}{2001}\natexlab{}.
\newblock \showarticletitle{Partitioning Digital Worlds: Focal and Peripheral
  Awareness in Multiple Monitor Use}. In \bibinfo{booktitle}{\emph{Proceedings
  of the SIGCHI Conference on Human Factors in Computing Systems}} (Seattle,
  Washington, USA) \emph{(\bibinfo{series}{CHI '01})}.
  \bibinfo{publisher}{Association for Computing Machinery},
  \bibinfo{address}{New York, NY, USA}, \bibinfo{pages}{458–465}.
\newblock
\showISBNx{1581133278}
\urldef\tempurl%
\url{https://doi.org/10.1145/365024.365312}
\showDOI{\tempurl}


\bibitem[Gutwin and Greenberg(2002)]%
        {gutwin2002descriptive}
\bibfield{author}{\bibinfo{person}{Carl Gutwin} {and} \bibinfo{person}{Saul
  Greenberg}.} \bibinfo{year}{2002}\natexlab{}.
\newblock \showarticletitle{A Descriptive Framework of Workspace Awareness for
  Real-Time Groupware}.
\newblock \bibinfo{journal}{\emph{Computer Supported Cooperative Work (CSCW)}}
  \bibinfo{volume}{11}, \bibinfo{number}{3} (\bibinfo{date}{01 Sep}
  \bibinfo{year}{2002}), \bibinfo{pages}{411--446}.
\newblock
\showISSN{1573-7551}
\urldef\tempurl%
\url{https://doi.org/10.1023/A:1021271517844}
\showDOI{\tempurl}


\bibitem[Hansen et~al\mbox{.}(2018)]%
        {Hansen2018GazeInteraction}
\bibfield{author}{\bibinfo{person}{John~Paulin Hansen}, \bibinfo{person}{Vijay
  Rajanna}, \bibinfo{person}{I.~Scott MacKenzie}, {and} \bibinfo{person}{Per
  B\ae{}kgaard}.} \bibinfo{year}{2018}\natexlab{}.
\newblock \showarticletitle{A Fitts' law study of click and dwell interaction
  by gaze, head and mouse with a head-mounted display}. In
  \bibinfo{booktitle}{\emph{Proceedings of the Workshop on Communication by
  Gaze Interaction}} (Warsaw, Poland) \emph{(\bibinfo{series}{COGAIN '18})}.
  \bibinfo{publisher}{Association for Computing Machinery},
  \bibinfo{address}{New York, NY, USA}, Article \bibinfo{articleno}{7},
  \bibinfo{numpages}{5}~pages.
\newblock
\showISBNx{9781450357906}
\urldef\tempurl%
\url{https://doi.org/10.1145/3206343.3206344}
\showDOI{\tempurl}


\bibitem[Hart(2006)]%
        {Sandra2006Nasa}
\bibfield{author}{\bibinfo{person}{Sandra~G. Hart}.}
  \bibinfo{year}{2006}\natexlab{}.
\newblock \showarticletitle{Nasa-Task Load Index (NASA-TLX); 20 Years Later}.
\newblock \bibinfo{journal}{\emph{Proceedings of the Human Factors and
  Ergonomics Society Annual Meeting}} \bibinfo{volume}{50}, \bibinfo{number}{9}
  (\bibinfo{year}{2006}), \bibinfo{pages}{904--908}.
\newblock
\urldef\tempurl%
\url{https://doi.org/10.1177/154193120605000909}
\showDOI{\tempurl}


\bibitem[Hoppe et~al\mbox{.}(2021)]%
        {Hoppe2021Shisha}
\bibfield{author}{\bibinfo{person}{Adrian~H. Hoppe}, \bibinfo{person}{Florian
  van~de Camp}, {and} \bibinfo{person}{Rainer Stiefelhagen}.}
  \bibinfo{year}{2021}\natexlab{}.
\newblock \showarticletitle{ShiSha: Enabling Shared Perspective With
  Face-to-Face Collaboration Using Redirected Avatars in Virtual Reality}.
\newblock \bibinfo{journal}{\emph{Proc. ACM Hum.-Comput. Interact.}}
  \bibinfo{volume}{4}, \bibinfo{number}{CSCW3}, Article
  \bibinfo{articleno}{251} (\bibinfo{date}{jan} \bibinfo{year}{2021}),
  \bibinfo{numpages}{22}~pages.
\newblock
\urldef\tempurl%
\url{https://doi.org/10.1145/3432950}
\showDOI{\tempurl}


\bibitem[Hutchings et~al\mbox{.}(2004)]%
        {Hutchings2004MultiDisplay}
\bibfield{author}{\bibinfo{person}{Dugald~Ralph Hutchings},
  \bibinfo{person}{Greg Smith}, \bibinfo{person}{Brian Meyers},
  \bibinfo{person}{Mary Czerwinski}, {and} \bibinfo{person}{George Robertson}.}
  \bibinfo{year}{2004}\natexlab{}.
\newblock \showarticletitle{Display Space Usage and Window Management Operation
  Comparisons between Single Monitor and Multiple Monitor Users}. In
  \bibinfo{booktitle}{\emph{Proceedings of the Working Conference on Advanced
  Visual Interfaces}} (Gallipoli, Italy) \emph{(\bibinfo{series}{AVI '04})}.
  \bibinfo{publisher}{Association for Computing Machinery},
  \bibinfo{address}{New York, NY, USA}, \bibinfo{pages}{32–39}.
\newblock
\showISBNx{1581138679}
\urldef\tempurl%
\url{https://doi.org/10.1145/989863.989867}
\showDOI{\tempurl}


\bibitem[Jetter et~al\mbox{.}(2011)]%
        {Jetter2011FaceStreams}
\bibfield{author}{\bibinfo{person}{Hans-Christian Jetter},
  \bibinfo{person}{Jens Gerken}, \bibinfo{person}{Michael Z\"{o}llner},
  \bibinfo{person}{Harald Reiterer}, {and} \bibinfo{person}{Natasa
  Milic-Frayling}.} \bibinfo{year}{2011}\natexlab{}.
\newblock \showarticletitle{Materializing the Query with Facet-Streams: A
  Hybrid Surface for Collaborative Search on Tabletops}. In
  \bibinfo{booktitle}{\emph{Proceedings of the SIGCHI Conference on Human
  Factors in Computing Systems}} (Vancouver, BC, Canada)
  \emph{(\bibinfo{series}{CHI '11})}. \bibinfo{publisher}{Association for
  Computing Machinery}, \bibinfo{address}{New York, NY, USA},
  \bibinfo{pages}{3013–3022}.
\newblock
\showISBNx{9781450302289}
\urldef\tempurl%
\url{https://doi.org/10.1145/1978942.1979390}
\showDOI{\tempurl}


\bibitem[Jing et~al\mbox{.}(2021)]%
        {Allison2021EyeMrVis}
\bibfield{author}{\bibinfo{person}{Allison Jing},
  \bibinfo{person}{Kieran~William May}, \bibinfo{person}{Mahnoor Naeem},
  \bibinfo{person}{Gun Lee}, {and} \bibinfo{person}{Mark Billinghurst}.}
  \bibinfo{year}{2021}\natexlab{}.
\newblock \showarticletitle{EyemR-Vis: Using Bi-Directional Gaze Behavioural
  Cues to Improve Mixed Reality Remote Collaboration}. In
  \bibinfo{booktitle}{\emph{Extended Abstracts of the 2021 CHI Conference on
  Human Factors in Computing Systems}} (Yokohama, Japan)
  \emph{(\bibinfo{series}{CHI EA '21})}. \bibinfo{publisher}{Association for
  Computing Machinery}, \bibinfo{address}{New York, NY, USA}, Article
  \bibinfo{articleno}{283}, \bibinfo{numpages}{7}~pages.
\newblock
\showISBNx{9781450380959}
\urldef\tempurl%
\url{https://doi.org/10.1145/3411763.3451844}
\showDOI{\tempurl}


\bibitem[Kang et~al\mbox{.}(2023)]%
        {Kang2023Retargeting}
\bibfield{author}{\bibinfo{person}{Jiho Kang}, \bibinfo{person}{Dongseok Yang},
  \bibinfo{person}{Taehei Kim}, \bibinfo{person}{Yewon Lee}, {and}
  \bibinfo{person}{Sung-Hee Lee}.} \bibinfo{year}{2023}\natexlab{}.
\newblock \showarticletitle{Real-time Retargeting of Deictic Motion to Virtual
  Avatars for Augmented Reality Telepresence}. In
  \bibinfo{booktitle}{\emph{2023 IEEE International Symposium on Mixed and
  Augmented Reality (ISMAR)}}. \bibinfo{pages}{885--893}.
\newblock
\urldef\tempurl%
\url{https://doi.org/10.1109/ISMAR59233.2023.00104}
\showDOI{\tempurl}


\bibitem[Kirk et~al\mbox{.}(2007)]%
        {kirk2007turn}
\bibfield{author}{\bibinfo{person}{David Kirk}, \bibinfo{person}{Tom Rodden},
  {and} \bibinfo{person}{Dana\"{e}~Stanton Fraser}.}
  \bibinfo{year}{2007}\natexlab{}.
\newblock \showarticletitle{Turn it this way: grounding collaborative action
  with remote gestures}. In \bibinfo{booktitle}{\emph{Proceedings of the SIGCHI
  Conference on Human Factors in Computing Systems}} (San Jose, California,
  USA) \emph{(\bibinfo{series}{CHI '07})}. \bibinfo{publisher}{Association for
  Computing Machinery}, \bibinfo{address}{New York, NY, USA},
  \bibinfo{pages}{1039–1048}.
\newblock
\showISBNx{9781595935939}
\urldef\tempurl%
\url{https://doi.org/10.1145/1240624.1240782}
\showDOI{\tempurl}


\bibitem[Kraut et~al\mbox{.}(2003)]%
        {kraut2003visual}
\bibfield{author}{\bibinfo{person}{Robert~E. Kraut}, \bibinfo{person}{Susan~R
  Fussell}, {and} \bibinfo{person}{Jane Siegel}.}
  \bibinfo{year}{2003}\natexlab{}.
\newblock \showarticletitle{Visual Information as a Conversational Resource in
  Collaborative Physical Tasks}.
\newblock \bibinfo{journal}{\emph{Human–Computer Interaction}}
  \bibinfo{volume}{18}, \bibinfo{number}{1-2} (\bibinfo{year}{2003}),
  \bibinfo{pages}{13--49}.
\newblock
\urldef\tempurl%
\url{https://doi.org/10.1207/S15327051HCI1812\_2}
\showDOI{\tempurl}


\bibitem[Lee et~al\mbox{.}(2021)]%
        {Lee2021Collaboration}
\bibfield{author}{\bibinfo{person}{Benjamin Lee}, \bibinfo{person}{Xiaoyun Hu},
  \bibinfo{person}{Maxime Cordeil}, \bibinfo{person}{Arnaud Prouzeau},
  \bibinfo{person}{Bernhard Jenny}, {and} \bibinfo{person}{Tim Dwyer}.}
  \bibinfo{year}{2021}\natexlab{}.
\newblock \showarticletitle{Shared Surfaces and Spaces: Collaborative Data
  Visualisation in a Co-located Immersive Environment}.
\newblock \bibinfo{journal}{\emph{IEEE Transactions on Visualization and
  Computer Graphics}} \bibinfo{volume}{27}, \bibinfo{number}{2}
  (\bibinfo{year}{2021}), \bibinfo{pages}{1171--1181}.
\newblock
\urldef\tempurl%
\url{https://doi.org/10.1109/TVCG.2020.3030450}
\showDOI{\tempurl}


\bibitem[Lehment et~al\mbox{.}(2014)]%
        {Lehment2014Alignment}
\bibfield{author}{\bibinfo{person}{Nicolas~H. Lehment}, \bibinfo{person}{Daniel
  Merget}, {and} \bibinfo{person}{Gerhard Rigoll}.}
  \bibinfo{year}{2014}\natexlab{}.
\newblock \showarticletitle{Creating automatically aligned consensus realities
  for AR videoconferencing}. In \bibinfo{booktitle}{\emph{2014 IEEE
  International Symposium on Mixed and Augmented Reality (ISMAR)}}.
  \bibinfo{pages}{201--206}.
\newblock
\urldef\tempurl%
\url{https://doi.org/10.1109/ISMAR.2014.6948428}
\showDOI{\tempurl}


\bibitem[Li et~al\mbox{.}(2019)]%
        {Zhen2019HoloDoc}
\bibfield{author}{\bibinfo{person}{Zhen Li}, \bibinfo{person}{Michelle Annett},
  \bibinfo{person}{Ken Hinckley}, \bibinfo{person}{Karan Singh}, {and}
  \bibinfo{person}{Daniel Wigdor}.} \bibinfo{year}{2019}\natexlab{}.
\newblock \showarticletitle{HoloDoc: Enabling Mixed Reality Workspaces That
  Harness Physical and Digital Content}. In
  \bibinfo{booktitle}{\emph{Proceedings of the 2019 CHI Conference on Human
  Factors in Computing Systems}} (Glasgow, Scotland Uk)
  \emph{(\bibinfo{series}{CHI '19})}. \bibinfo{publisher}{Association for
  Computing Machinery}, \bibinfo{address}{New York, NY, USA},
  \bibinfo{pages}{1–14}.
\newblock
\showISBNx{9781450359702}
\urldef\tempurl%
\url{https://doi.org/10.1145/3290605.3300917}
\showDOI{\tempurl}


\bibitem[Lindlbauer(2022)]%
        {Lindlbauer2022Adaptive}
\bibfield{author}{\bibinfo{person}{David Lindlbauer}.}
  \bibinfo{year}{2022}\natexlab{}.
\newblock \showarticletitle{The Future of Mixed Reality is Adaptive}.
\newblock \bibinfo{journal}{\emph{XRDS}} \bibinfo{volume}{29},
  \bibinfo{number}{1} (\bibinfo{date}{oct} \bibinfo{year}{2022}),
  \bibinfo{pages}{26–31}.
\newblock
\showISSN{1528-4972}
\urldef\tempurl%
\url{https://doi.org/10.1145/3558191}
\showDOI{\tempurl}


\bibitem[Lindlbauer et~al\mbox{.}(2019)]%
        {Lindlbauer19}
\bibfield{author}{\bibinfo{person}{David Lindlbauer},
  \bibinfo{person}{Anna~Maria Feit}, {and} \bibinfo{person}{Otmar Hilliges}.}
  \bibinfo{year}{2019}\natexlab{}.
\newblock \showarticletitle{Context-aware online adaptation of mixed reality
  interfaces}.
\newblock \bibinfo{journal}{\emph{UIST 2019 - Proceedings of the 32nd Annual
  ACM Symposium on User Interface Software and Technology}},
  \bibinfo{pages}{147--160}.
\newblock
\showISBNx{9781450368162}
\urldef\tempurl%
\url{https://doi.org/10.1145/3332165.3347945}
\showDOI{\tempurl}


\bibitem[Ling et~al\mbox{.}(2017)]%
        {Ling2017Workspace}
\bibfield{author}{\bibinfo{person}{Chen Ling}, \bibinfo{person}{Alex Stegman},
  \bibinfo{person}{Chintan Barhbaya}, {and} \bibinfo{person}{Randa Shehab}.}
  \bibinfo{year}{2017}\natexlab{}.
\newblock \showarticletitle{Are Two Better Than One? A Comparison Between
  Single- and Dual-Monitor Work Stations in Productivity and User’s Windows
  Management Style}.
\newblock \bibinfo{journal}{\emph{International Journal of Human–Computer
  Interaction}} \bibinfo{volume}{33}, \bibinfo{number}{3}
  (\bibinfo{year}{2017}), \bibinfo{pages}{190--198}.
\newblock
\urldef\tempurl%
\url{https://doi.org/10.1080/10447318.2016.1231392}
\showDOI{\tempurl}


\bibitem[Lischke et~al\mbox{.}(2016)]%
        {Lischke16}
\bibfield{author}{\bibinfo{person}{Lars Lischke}, \bibinfo{person}{Sven Mayer},
  \bibinfo{person}{Katrin Wolf}, \bibinfo{person}{Niels Henze},
  \bibinfo{person}{Harald Reiterer}, {and} \bibinfo{person}{Albrecht Schmidt}.}
  \bibinfo{year}{2016}\natexlab{}.
\newblock \showarticletitle{Screen arrangements and interaction areas for large
  display work places}.
\newblock \bibinfo{journal}{\emph{PerDis 2016 - Proceedings of the 5th ACM
  International Symposium on Pervasive Displays}}, \bibinfo{pages}{228--234}.
\newblock
\showISBNx{9781450343664}
\urldef\tempurl%
\url{https://doi.org/10.1145/2914920.2915027}
\showDOI{\tempurl}


\bibitem[Lombard et~al\mbox{.}(2009)]%
        {Lombard2009tpi}
\bibfield{author}{\bibinfo{person}{Matthew Lombard}, \bibinfo{person}{Theresa~B
  Ditton}, {and} \bibinfo{person}{Lisa Weinstein}.}
  \bibinfo{year}{2009}\natexlab{}.
\newblock \showarticletitle{Measuring presence: the temple presence inventory}.
  In \bibinfo{booktitle}{\emph{Proceedings of the 12th annual international
  workshop on presence}}. \bibinfo{pages}{1--15}.
\newblock


\bibitem[Luo et~al\mbox{.}(2022)]%
        {Luo22Collaboration}
\bibfield{author}{\bibinfo{person}{Weizhou Luo}, \bibinfo{person}{Anke
  Lehmann}, \bibinfo{person}{Hjalmar Widengren}, {and} \bibinfo{person}{Raimund
  Dachselt}.} \bibinfo{year}{2022}\natexlab{}.
\newblock \showarticletitle{Where Should We Put It? Layout and Placement
  Strategies of Documents in Augmented Reality for Collaborative Sensemaking}.
  In \bibinfo{booktitle}{\emph{Proceedings of the 2022 CHI Conference on Human
  Factors in Computing Systems}} (New Orleans, LA, USA)
  \emph{(\bibinfo{series}{CHI '22})}. \bibinfo{publisher}{Association for
  Computing Machinery}, \bibinfo{address}{New York, NY, USA}, Article
  \bibinfo{articleno}{627}, \bibinfo{numpages}{16}~pages.
\newblock
\showISBNx{9781450391573}
\urldef\tempurl%
\url{https://doi.org/10.1145/3491102.3501946}
\showDOI{\tempurl}


\bibitem[Marquardt et~al\mbox{.}(2012)]%
        {Marquardt2012CrossDevice}
\bibfield{author}{\bibinfo{person}{Nicolai Marquardt}, \bibinfo{person}{Ken
  Hinckley}, {and} \bibinfo{person}{Saul Greenberg}.}
  \bibinfo{year}{2012}\natexlab{}.
\newblock \showarticletitle{Cross-Device Interaction via Micro-Mobility and
  f-Formations}. In \bibinfo{booktitle}{\emph{Proceedings of the 25th Annual
  ACM Symposium on User Interface Software and Technology}} (Cambridge,
  Massachusetts, USA) \emph{(\bibinfo{series}{UIST '12})}.
  \bibinfo{publisher}{Association for Computing Machinery},
  \bibinfo{address}{New York, NY, USA}, \bibinfo{pages}{13–22}.
\newblock
\showISBNx{9781450315807}
\urldef\tempurl%
\url{https://doi.org/10.1145/2380116.2380121}
\showDOI{\tempurl}


\bibitem[Mcgill et~al\mbox{.}(2020)]%
        {Mcgill2020Workspaces}
\bibfield{author}{\bibinfo{person}{Mark Mcgill}, \bibinfo{person}{Aidan Kehoe},
  \bibinfo{person}{Euan Freeman}, {and} \bibinfo{person}{Stephen Brewster}.}
  \bibinfo{year}{2020}\natexlab{}.
\newblock \showarticletitle{Expanding the Bounds of Seated Virtual Workspaces}.
\newblock \bibinfo{journal}{\emph{ACM Trans. Comput.-Hum. Interact.}}
  \bibinfo{volume}{27}, \bibinfo{number}{3}, Article \bibinfo{articleno}{13}
  (\bibinfo{date}{may} \bibinfo{year}{2020}), \bibinfo{numpages}{40}~pages.
\newblock
\showISSN{1073-0516}
\urldef\tempurl%
\url{https://doi.org/10.1145/3380959}
\showDOI{\tempurl}


\bibitem[Medeiros et~al\mbox{.}(2022)]%
        {Medeiros2022PassengerWorkspace}
\bibfield{author}{\bibinfo{person}{Daniel Medeiros}, \bibinfo{person}{Mark
  McGill}, \bibinfo{person}{Alexander Ng}, \bibinfo{person}{Robert McDermid},
  \bibinfo{person}{Nadia Pantidi}, \bibinfo{person}{Julie Williamson}, {and}
  \bibinfo{person}{Stephen Brewster}.} \bibinfo{year}{2022}\natexlab{}.
\newblock \showarticletitle{From Shielding to Avoidance: Passenger Augmented
  Reality and the Layout of Virtual Displays for Productivity in Shared
  Transit}.
\newblock \bibinfo{journal}{\emph{IEEE Transactions on Visualization and
  Computer Graphics}} \bibinfo{volume}{28}, \bibinfo{number}{11}
  (\bibinfo{year}{2022}), \bibinfo{pages}{3640--3650}.
\newblock
\urldef\tempurl%
\url{https://doi.org/10.1109/TVCG.2022.3203002}
\showDOI{\tempurl}


\bibitem[Morris(2008)]%
        {Morris2008WebSearch}
\bibfield{author}{\bibinfo{person}{Meredith~Ringel Morris}.}
  \bibinfo{year}{2008}\natexlab{}.
\newblock \showarticletitle{A Survey of Collaborative Web Search Practices}. In
  \bibinfo{booktitle}{\emph{Proceedings of the SIGCHI Conference on Human
  Factors in Computing Systems}} (Florence, Italy) \emph{(\bibinfo{series}{CHI
  '08})}. \bibinfo{publisher}{Association for Computing Machinery},
  \bibinfo{address}{New York, NY, USA}, \bibinfo{pages}{1657–1660}.
\newblock
\showISBNx{9781605580111}
\urldef\tempurl%
\url{https://doi.org/10.1145/1357054.1357312}
\showDOI{\tempurl}


\bibitem[Ng et~al\mbox{.}(2021)]%
        {Ng2021PassengerWorkspace}
\bibfield{author}{\bibinfo{person}{Alexander Ng}, \bibinfo{person}{Daniel
  Medeiros}, \bibinfo{person}{Mark McGill}, \bibinfo{person}{Julie Williamson},
  {and} \bibinfo{person}{Stephen Brewster}.} \bibinfo{year}{2021}\natexlab{}.
\newblock \showarticletitle{The Passenger Experience of Mixed Reality Virtual
  Display Layouts in Airplane Environments}. In \bibinfo{booktitle}{\emph{2021
  IEEE International Symposium on Mixed and Augmented Reality (ISMAR)}}.
  \bibinfo{pages}{265--274}.
\newblock
\urldef\tempurl%
\url{https://doi.org/10.1109/ISMAR52148.2021.00042}
\showDOI{\tempurl}


\bibitem[Nguyen and Canny(2005)]%
        {Nguyen2005MultiView}
\bibfield{author}{\bibinfo{person}{David Nguyen} {and} \bibinfo{person}{John
  Canny}.} \bibinfo{year}{2005}\natexlab{}.
\newblock \showarticletitle{MultiView: Spatially Faithful Group Video
  Conferencing}. In \bibinfo{booktitle}{\emph{Proceedings of the SIGCHI
  Conference on Human Factors in Computing Systems}} (Portland, Oregon, USA)
  \emph{(\bibinfo{series}{CHI '05})}. \bibinfo{publisher}{Association for
  Computing Machinery}, \bibinfo{address}{New York, NY, USA},
  \bibinfo{pages}{799–808}.
\newblock
\showISBNx{1581139985}
\urldef\tempurl%
\url{https://doi.org/10.1145/1054972.1055084}
\showDOI{\tempurl}


\bibitem[Niyazov et~al\mbox{.}(2023)]%
        {Niyazov23}
\bibfield{author}{\bibinfo{person}{Aziz Niyazov}, \bibinfo{person}{Barrett
  Ens}, \bibinfo{person}{Kadek~Ananta Satriadi}, \bibinfo{person}{Nicolas
  Mellado}, \bibinfo{person}{Loic Barthe}, \bibinfo{person}{Tim Dwyer}, {and}
  \bibinfo{person}{Marcos Serrano}.} \bibinfo{year}{2023}\natexlab{}.
\newblock \showarticletitle{User-Driven Constraints for Layout Optimisation in
  Augmented Reality}.
\newblock \bibinfo{journal}{\emph{Proceedings of the 2023 CHI Conference on
  Human Factors in Computing Systems}}, \bibinfo{pages}{1--16}.
\newblock
\showISBNx{9781450394215}
\urldef\tempurl%
\url{https://doi.org/10.1145/3544548.3580873}
\showDOI{\tempurl}


\bibitem[O'hara et~al\mbox{.}(2011)]%
        {Ohara2011Blended}
\bibfield{author}{\bibinfo{person}{Kenton O'hara}, \bibinfo{person}{Jesper
  Kjeldskov}, {and} \bibinfo{person}{Jeni Paay}.}
  \bibinfo{year}{2011}\natexlab{}.
\newblock \showarticletitle{Blended Interaction Spaces for Distributed Team
  Collaboration}.
\newblock \bibinfo{journal}{\emph{ACM Trans. Comput.-Hum. Interact.}}
  \bibinfo{volume}{18}, \bibinfo{number}{1}, Article \bibinfo{articleno}{3}
  (\bibinfo{date}{may} \bibinfo{year}{2011}), \bibinfo{numpages}{28}~pages.
\newblock
\showISSN{1073-0516}
\urldef\tempurl%
\url{https://doi.org/10.1145/1959022.1959025}
\showDOI{\tempurl}


\bibitem[Owens et~al\mbox{.}(2012)]%
        {Owens2012Workspace}
\bibfield{author}{\bibinfo{person}{Justin~W. Owens}, \bibinfo{person}{Jennifer
  Teves}, \bibinfo{person}{Bobby Nguyen}, \bibinfo{person}{Amanda Smith},
  \bibinfo{person}{Mandy~C. Phelps}, {and} \bibinfo{person}{Barbara~S.
  Chaparro}.} \bibinfo{year}{2012}\natexlab{}.
\newblock \showarticletitle{Examination of Dual vs. Single Monitor Use during
  Common Office Tasks}.
\newblock \bibinfo{journal}{\emph{Proceedings of the Human Factors and
  Ergonomics Society Annual Meeting}} \bibinfo{volume}{56}, \bibinfo{number}{1}
  (\bibinfo{year}{2012}), \bibinfo{pages}{1506--1510}.
\newblock
\urldef\tempurl%
\url{https://doi.org/10.1177/1071181312561299}
\showDOI{\tempurl}


\bibitem[Pavanatto(2021)]%
        {Pavanatto2021VirtualScreens}
\bibfield{author}{\bibinfo{person}{Leonardo Pavanatto}.}
  \bibinfo{year}{2021}\natexlab{}.
\newblock \showarticletitle{Designing Augmented Reality Virtual Displays for
  Productivity Work}. In \bibinfo{booktitle}{\emph{2021 IEEE International
  Symposium on Mixed and Augmented Reality Adjunct (ISMAR-Adjunct)}}.
  \bibinfo{pages}{459--460}.
\newblock
\urldef\tempurl%
\url{https://doi.org/10.1109/ISMAR-Adjunct54149.2021.00107}
\showDOI{\tempurl}


\bibitem[Pavanatto et~al\mbox{.}(2021)]%
        {Pavanatto21}
\bibfield{author}{\bibinfo{person}{Leonardo Pavanatto}, \bibinfo{person}{Chris
  North}, \bibinfo{person}{Doug~A. Bowman}, \bibinfo{person}{Carmen Badea},
  {and} \bibinfo{person}{Richard Stoakley}.} \bibinfo{year}{2021}\natexlab{}.
\newblock \showarticletitle{Do we still need physical monitors? An evaluation
  of the usability of AR virtual monitors for productivity work}. In
  \bibinfo{booktitle}{\emph{2021 IEEE Virtual Reality and 3D User Interfaces
  (VR)}} (Lisboa, Portugal). \bibinfo{publisher}{IEEE},
  \bibinfo{pages}{759--767}.
\newblock
\urldef\tempurl%
\url{https://doi.org/10.1109/VR50410.2021.00103}
\showDOI{\tempurl}


\bibitem[Pejsa et~al\mbox{.}(2016)]%
        {Pejsa2016Room}
\bibfield{author}{\bibinfo{person}{Tomislav Pejsa}, \bibinfo{person}{Julian
  Kantor}, \bibinfo{person}{Hrvoje Benko}, \bibinfo{person}{Eyal Ofek}, {and}
  \bibinfo{person}{Andrew Wilson}.} \bibinfo{year}{2016}\natexlab{}.
\newblock \showarticletitle{Room2Room: Enabling Life-Size Telepresence in a
  Projected Augmented Reality Environment}. In
  \bibinfo{booktitle}{\emph{Proceedings of the 19th ACM Conference on
  Computer-Supported Cooperative Work \& Social Computing}} (San Francisco,
  California, USA) \emph{(\bibinfo{series}{CSCW '16})}.
  \bibinfo{publisher}{Association for Computing Machinery},
  \bibinfo{address}{New York, NY, USA}, \bibinfo{pages}{1716–1725}.
\newblock
\showISBNx{9781450335928}
\urldef\tempurl%
\url{https://doi.org/10.1145/2818048.2819965}
\showDOI{\tempurl}


\bibitem[Piumsomboon et~al\mbox{.}(2019)]%
        {Piumsomboon19}
\bibfield{author}{\bibinfo{person}{Thammathip Piumsomboon},
  \bibinfo{person}{Arindam Dey}, \bibinfo{person}{Barrett Ens},
  \bibinfo{person}{Gun Lee}, {and} \bibinfo{person}{Mark Billinghurst}.}
  \bibinfo{year}{2019}\natexlab{}.
\newblock \showarticletitle{The effects of sharing awareness cues in
  collaborative mixed reality}.
\newblock \bibinfo{journal}{\emph{Frontiers Robotics AI}}  \bibinfo{volume}{6}
  (\bibinfo{year}{2019}).
\newblock
Issue FEB.
\showISSN{22969144}
\urldef\tempurl%
\url{https://doi.org/10.3389/frobt.2019.00005}
\showDOI{\tempurl}


\bibitem[Piumsomboon et~al\mbox{.}(2018)]%
        {Piumsomboon2018MiniMe}
\bibfield{author}{\bibinfo{person}{Thammathip Piumsomboon},
  \bibinfo{person}{Gun~A. Lee}, \bibinfo{person}{Jonathon~D. Hart},
  \bibinfo{person}{Barrett Ens}, \bibinfo{person}{Robert~W. Lindeman},
  \bibinfo{person}{Bruce~H. Thomas}, {and} \bibinfo{person}{Mark
  Billinghurst}.} \bibinfo{year}{2018}\natexlab{}.
\newblock \showarticletitle{Mini-Me: An Adaptive Avatar for Mixed Reality
  Remote Collaboration}. In \bibinfo{booktitle}{\emph{Proceedings of the 2018
  CHI Conference on Human Factors in Computing Systems}} (Montreal QC, Canada)
  \emph{(\bibinfo{series}{CHI '18})}. \bibinfo{publisher}{Association for
  Computing Machinery}, \bibinfo{address}{New York, NY, USA},
  \bibinfo{pages}{1–13}.
\newblock
\showISBNx{9781450356206}
\urldef\tempurl%
\url{https://doi.org/10.1145/3173574.3173620}
\showDOI{\tempurl}


\bibitem[Plasson et~al\mbox{.}(2022)]%
        {Plasson2022DesktopAR}
\bibfield{author}{\bibinfo{person}{Carole Plasson}, \bibinfo{person}{Renaud
  Blanch}, {and} \bibinfo{person}{Laurence Nigay}.}
  \bibinfo{year}{2022}\natexlab{}.
\newblock \showarticletitle{Selection Techniques for 3D Extended Desktop
  Workstation with AR HMD}. In \bibinfo{booktitle}{\emph{2022 IEEE
  International Symposium on Mixed and Augmented Reality (ISMAR)}}.
  \bibinfo{pages}{460--469}.
\newblock
\urldef\tempurl%
\url{https://doi.org/10.1109/ISMAR55827.2022.00062}
\showDOI{\tempurl}


\bibitem[Qian et~al\mbox{.}(2022b)]%
        {Qian2022DuallyNoted}
\bibfield{author}{\bibinfo{person}{Jing Qian}, \bibinfo{person}{Qi Sun},
  \bibinfo{person}{Curtis Wigington}, \bibinfo{person}{Han~L. Han},
  \bibinfo{person}{Tong Sun}, \bibinfo{person}{Jennifer Healey},
  \bibinfo{person}{James Tompkin}, {and} \bibinfo{person}{Jeff Huang}.}
  \bibinfo{year}{2022}\natexlab{b}.
\newblock \showarticletitle{Dually Noted: Layout-Aware Annotations with
  Smartphone Augmented Reality}. In \bibinfo{booktitle}{\emph{Proceedings of
  the 2022 CHI Conference on Human Factors in Computing Systems}} (New Orleans,
  LA, USA) \emph{(\bibinfo{series}{CHI '22})}. \bibinfo{publisher}{Association
  for Computing Machinery}, \bibinfo{address}{New York, NY, USA}, Article
  \bibinfo{articleno}{552}, \bibinfo{numpages}{15}~pages.
\newblock
\showISBNx{9781450391573}
\urldef\tempurl%
\url{https://doi.org/10.1145/3491102.3502026}
\showDOI{\tempurl}


\bibitem[Qian et~al\mbox{.}(2022a)]%
        {Qian2022Scalar}
\bibfield{author}{\bibinfo{person}{Xun Qian}, \bibinfo{person}{Fengming He},
  \bibinfo{person}{Xiyun Hu}, \bibinfo{person}{Tianyi Wang},
  \bibinfo{person}{Ananya Ipsita}, {and} \bibinfo{person}{Karthik Ramani}.}
  \bibinfo{year}{2022}\natexlab{a}.
\newblock \showarticletitle{ScalAR: Authoring Semantically Adaptive Augmented
  Reality Experiences in Virtual Reality}. In
  \bibinfo{booktitle}{\emph{Proceedings of the 2022 CHI Conference on Human
  Factors in Computing Systems}} (New Orleans, LA, USA)
  \emph{(\bibinfo{series}{CHI '22})}. \bibinfo{publisher}{Association for
  Computing Machinery}, \bibinfo{address}{New York, NY, USA}, Article
  \bibinfo{articleno}{65}, \bibinfo{numpages}{18}~pages.
\newblock
\showISBNx{9781450391573}
\urldef\tempurl%
\url{https://doi.org/10.1145/3491102.3517665}
\showDOI{\tempurl}


\bibitem[Rajaram and Nebeling(2022)]%
        {Rajaram2022PaperTrail}
\bibfield{author}{\bibinfo{person}{Shwetha Rajaram} {and}
  \bibinfo{person}{Michael Nebeling}.} \bibinfo{year}{2022}\natexlab{}.
\newblock \showarticletitle{Paper Trail: An Immersive Authoring System for
  Augmented Reality Instructional Experiences}. In
  \bibinfo{booktitle}{\emph{Proceedings of the 2022 CHI Conference on Human
  Factors in Computing Systems}} (New Orleans, LA, USA)
  \emph{(\bibinfo{series}{CHI '22})}. \bibinfo{publisher}{Association for
  Computing Machinery}, \bibinfo{address}{New York, NY, USA}, Article
  \bibinfo{articleno}{382}, \bibinfo{numpages}{16}~pages.
\newblock
\showISBNx{9781450391573}
\urldef\tempurl%
\url{https://doi.org/10.1145/3491102.3517486}
\showDOI{\tempurl}


\bibitem[Satriadi et~al\mbox{.}(2020)]%
        {Satriadi2020Maps}
\bibfield{author}{\bibinfo{person}{Kadek~Ananta Satriadi},
  \bibinfo{person}{Barrett Ens}, \bibinfo{person}{Maxime Cordeil},
  \bibinfo{person}{Tobias Czauderna}, {and} \bibinfo{person}{Bernhard Jenny}.}
  \bibinfo{year}{2020}\natexlab{}.
\newblock \showarticletitle{Maps Around Me: 3D Multiview Layouts in Immersive
  Spaces}.
\newblock \bibinfo{journal}{\emph{Proc. ACM Hum.-Comput. Interact.}}
  \bibinfo{volume}{4}, \bibinfo{number}{ISS}, Article \bibinfo{articleno}{201}
  (\bibinfo{date}{nov} \bibinfo{year}{2020}), \bibinfo{numpages}{20}~pages.
\newblock
\urldef\tempurl%
\url{https://doi.org/10.1145/3427329}
\showDOI{\tempurl}


\bibitem[Schneider and Pea(2013)]%
        {Schneider2013Collaboration}
\bibfield{author}{\bibinfo{person}{Bertrand Schneider} {and}
  \bibinfo{person}{Roy Pea}.} \bibinfo{year}{2013}\natexlab{}.
\newblock \showarticletitle{Real-time mutual gaze perception enhances
  collaborative learning and collaboration quality}.
\newblock \bibinfo{journal}{\emph{International Journal of Computer-Supported
  Collaborative Learning}} \bibinfo{volume}{8}, \bibinfo{number}{4}
  (\bibinfo{date}{01 Dec} \bibinfo{year}{2013}), \bibinfo{pages}{375--397}.
\newblock
\showISSN{1556-1615}
\urldef\tempurl%
\url{https://doi.org/10.1007/s11412-013-9181-4}
\showDOI{\tempurl}


\bibitem[Shockley et~al\mbox{.}(2009)]%
        {shockley2009conversation}
\bibfield{author}{\bibinfo{person}{Kevin Shockley}, \bibinfo{person}{Daniel~C.
  Richardson}, {and} \bibinfo{person}{Rick Dale}.}
  \bibinfo{year}{2009}\natexlab{}.
\newblock \showarticletitle{Conversation and Coordinative Structures}.
\newblock \bibinfo{journal}{\emph{Topics in Cognitive Science}}
  \bibinfo{volume}{1}, \bibinfo{number}{2} (\bibinfo{year}{2009}),
  \bibinfo{pages}{305--319}.
\newblock
\urldef\tempurl%
\url{https://doi.org/10.1111/j.1756-8765.2009.01021.x}
\showDOI{\tempurl}


\bibitem[Sousa et~al\mbox{.}(2019)]%
        {Sousa19}
\bibfield{author}{\bibinfo{person}{Maur\'{\i}cio Sousa},
  \bibinfo{person}{Rafael~Kufner dos Anjos}, \bibinfo{person}{Daniel Mendes},
  \bibinfo{person}{Mark Billinghurst}, {and} \bibinfo{person}{Joaquim Jorge}.}
  \bibinfo{year}{2019}\natexlab{}.
\newblock \showarticletitle{Warping Deixis: Distorting Gestures to Enhance
  Collaboration}. In \bibinfo{booktitle}{\emph{Proceedings of the 2019 CHI
  Conference on Human Factors in Computing Systems}} (Glasgow, Scotland Uk)
  \emph{(\bibinfo{series}{CHI '19})}. \bibinfo{publisher}{Association for
  Computing Machinery}, \bibinfo{address}{New York, NY, USA},
  \bibinfo{pages}{1–12}.
\newblock
\showISBNx{9781450359702}
\urldef\tempurl%
\url{https://doi.org/10.1145/3290605.3300838}
\showDOI{\tempurl}


\bibitem[Tan et~al\mbox{.}(2006)]%
        {Tan2006Workspace}
\bibfield{author}{\bibinfo{person}{Desney~S. Tan}, \bibinfo{person}{Darren
  Gergle}, \bibinfo{person}{Peter Scupelli}, {and} \bibinfo{person}{Randy
  Pausch}.} \bibinfo{year}{2006}\natexlab{}.
\newblock \showarticletitle{Physically Large Displays Improve Performance on
  Spatial Tasks}.
\newblock \bibinfo{journal}{\emph{ACM Trans. Comput.-Hum. Interact.}}
  \bibinfo{volume}{13}, \bibinfo{number}{1} (\bibinfo{date}{mar}
  \bibinfo{year}{2006}), \bibinfo{pages}{71–99}.
\newblock
\showISSN{1073-0516}
\urldef\tempurl%
\url{https://doi.org/10.1145/1143518.1143521}
\showDOI{\tempurl}


\bibitem[Thoravi~Kumaravel et~al\mbox{.}(2019)]%
        {Thoravi2019Loki}
\bibfield{author}{\bibinfo{person}{Balasaravanan Thoravi~Kumaravel},
  \bibinfo{person}{Fraser Anderson}, \bibinfo{person}{George Fitzmaurice},
  \bibinfo{person}{Bjoern Hartmann}, {and} \bibinfo{person}{Tovi Grossman}.}
  \bibinfo{year}{2019}\natexlab{}.
\newblock \showarticletitle{Loki: Facilitating Remote Instruction of Physical
  Tasks Using Bi-Directional Mixed-Reality Telepresence}. In
  \bibinfo{booktitle}{\emph{Proceedings of the 32nd Annual ACM Symposium on
  User Interface Software and Technology}} (New Orleans, LA, USA)
  \emph{(\bibinfo{series}{UIST '19})}. \bibinfo{publisher}{Association for
  Computing Machinery}, \bibinfo{address}{New York, NY, USA},
  \bibinfo{pages}{161–174}.
\newblock
\showISBNx{9781450368162}
\urldef\tempurl%
\url{https://doi.org/10.1145/3332165.3347872}
\showDOI{\tempurl}


\bibitem[Vertegaal et~al\mbox{.}(2003)]%
        {Vertegaal2003Gaze2}
\bibfield{author}{\bibinfo{person}{Roel Vertegaal}, \bibinfo{person}{Ivo
  Weevers}, \bibinfo{person}{Changuk Sohn}, {and} \bibinfo{person}{Chris
  Cheung}.} \bibinfo{year}{2003}\natexlab{}.
\newblock \showarticletitle{GAZE-2: Conveying Eye Contact in Group Video
  Conferencing Using Eye-Controlled Camera Direction}. In
  \bibinfo{booktitle}{\emph{Proceedings of the SIGCHI Conference on Human
  Factors in Computing Systems}} (Ft. Lauderdale, Florida, USA)
  \emph{(\bibinfo{series}{CHI '03})}. \bibinfo{publisher}{Association for
  Computing Machinery}, \bibinfo{address}{New York, NY, USA},
  \bibinfo{pages}{521–528}.
\newblock
\showISBNx{1581136307}
\urldef\tempurl%
\url{https://doi.org/10.1145/642611.642702}
\showDOI{\tempurl}


\bibitem[Wang et~al\mbox{.}(2021)]%
        {Wang2021Review}
\bibfield{author}{\bibinfo{person}{Peng Wang}, \bibinfo{person}{Xiaoliang Bai},
  \bibinfo{person}{Mark Billinghurst}, \bibinfo{person}{Shusheng Zhang},
  \bibinfo{person}{Xiangyu Zhang}, \bibinfo{person}{Shuxia Wang},
  \bibinfo{person}{Weiping He}, \bibinfo{person}{Yuxiang Yan}, {and}
  \bibinfo{person}{Hongyu Ji}.} \bibinfo{year}{2021}\natexlab{}.
\newblock \showarticletitle{AR/MR Remote Collaboration on Physical Tasks: A
  Review}.
\newblock \bibinfo{journal}{\emph{Robotics and Computer-Integrated
  Manufacturing}}  \bibinfo{volume}{72} (\bibinfo{year}{2021}),
  \bibinfo{pages}{102071}.
\newblock
\showISSN{0736-5845}
\urldef\tempurl%
\url{https://doi.org/10.1016/j.rcim.2020.102071}
\showDOI{\tempurl}


\bibitem[Williams and Kessler(2002)]%
        {Williams2002Programming}
\bibfield{author}{\bibinfo{person}{Laurie Williams} {and}
  \bibinfo{person}{Robert Kessler}.} \bibinfo{year}{2002}\natexlab{}.
\newblock \bibinfo{booktitle}{\emph{Pair Programming Illuminated}}.
\newblock \bibinfo{publisher}{Addison-Wesley Longman Publishing Co., Inc.},
  \bibinfo{address}{USA}.
\newblock
\showISBNx{0201745763}


\bibitem[Williamson et~al\mbox{.}(2019)]%
        {Williamson2019PlaneVr}
\bibfield{author}{\bibinfo{person}{Julie~R. Williamson}, \bibinfo{person}{Mark
  McGill}, {and} \bibinfo{person}{Khari Outram}.}
  \bibinfo{year}{2019}\natexlab{}.
\newblock \showarticletitle{PlaneVR: Social Acceptability of Virtual Reality
  for Aeroplane Passengers}. In \bibinfo{booktitle}{\emph{Proceedings of the
  2019 CHI Conference on Human Factors in Computing Systems}} (Glasgow,
  Scotland UK) \emph{(\bibinfo{series}{CHI '19})}.
  \bibinfo{publisher}{Association for Computing Machinery},
  \bibinfo{address}{New York, NY, USA}, \bibinfo{pages}{1–14}.
\newblock
\showISBNx{9781450359702}
\urldef\tempurl%
\url{https://doi.org/10.1145/3290605.3300310}
\showDOI{\tempurl}


\bibitem[Wong and Gutwin(2014)]%
        {Wong2014Gestures}
\bibfield{author}{\bibinfo{person}{Nelson Wong} {and} \bibinfo{person}{Carl
  Gutwin}.} \bibinfo{year}{2014}\natexlab{}.
\newblock \showarticletitle{Support for Deictic Pointing in CVEs: Still
  Fragmented after All These Years'}. In \bibinfo{booktitle}{\emph{Proceedings
  of the 17th ACM Conference on Computer Supported Cooperative Work \& Social
  Computing}} (Baltimore, Maryland, USA) \emph{(\bibinfo{series}{CSCW '14})}.
  \bibinfo{publisher}{Association for Computing Machinery},
  \bibinfo{address}{New York, NY, USA}, \bibinfo{pages}{1377–1387}.
\newblock
\showISBNx{9781450325400}
\urldef\tempurl%
\url{https://doi.org/10.1145/2531602.2531691}
\showDOI{\tempurl}


\bibitem[Yoon et~al\mbox{.}(2021)]%
        {Yoon21}
\bibfield{author}{\bibinfo{person}{Leonard Yoon}, \bibinfo{person}{Dongseok
  Yang}, \bibinfo{person}{Choongho Chung}, {and} \bibinfo{person}{Sung{-}Hee
  Lee}.} \bibinfo{year}{2021}\natexlab{}.
\newblock \showarticletitle{A Full Body Avatar-Based Telepresence System for
  Dissimilar Spaces}.
\newblock \bibinfo{journal}{\emph{CoRR}}  \bibinfo{volume}{abs/2103.04380}
  (\bibinfo{year}{2021}).
\newblock
\showeprint[arXiv]{2103.04380}
\urldef\tempurl%
\url{https://arxiv.org/abs/2103.04380}
\showURL{%
\tempurl}


\bibitem[Yuan et~al\mbox{.}(2022)]%
        {yuan2022understanding}
\bibfield{author}{\bibinfo{person}{Ye Yuan}, \bibinfo{person}{Nathalie Riche},
  \bibinfo{person}{Nicolai Marquardt}, \bibinfo{person}{Molly~Jane Nicholas},
  \bibinfo{person}{Teddy Seyed}, \bibinfo{person}{Hugo Romat},
  \bibinfo{person}{Bongshin Lee}, \bibinfo{person}{Michel Pahud},
  \bibinfo{person}{Jonathan Goldstein}, \bibinfo{person}{Rojin Vishkaie},
  \bibinfo{person}{Christian Holz}, {and} \bibinfo{person}{Ken Hinckley}.}
  \bibinfo{year}{2022}\natexlab{}.
\newblock \showarticletitle{Understanding Multi-Device Usage Patterns: Physical
  Device Configurations and Fragmented Workflows}. In
  \bibinfo{booktitle}{\emph{Proceedings of the 2022 CHI Conference on Human
  Factors in Computing Systems}} (New Orleans, LA, USA)
  \emph{(\bibinfo{series}{CHI '22})}. \bibinfo{publisher}{Association for
  Computing Machinery}, \bibinfo{address}{New York, NY, USA}, Article
  \bibinfo{articleno}{64}, \bibinfo{numpages}{22}~pages.
\newblock
\showISBNx{9781450391573}
\urldef\tempurl%
\url{https://doi.org/10.1145/3491102.3517702}
\showDOI{\tempurl}


\bibitem[Zhao et~al\mbox{.}(2007)]%
        {Zhao2007Earpod}
\bibfield{author}{\bibinfo{person}{Shengdong Zhao}, \bibinfo{person}{Pierre
  Dragicevic}, \bibinfo{person}{Mark Chignell}, \bibinfo{person}{Ravin
  Balakrishnan}, {and} \bibinfo{person}{Patrick Baudisch}.}
  \bibinfo{year}{2007}\natexlab{}.
\newblock \showarticletitle{Earpod: eyes-free menu selection using touch input
  and reactive audio feedback}. In \bibinfo{booktitle}{\emph{Proceedings of the
  SIGCHI Conference on Human Factors in Computing Systems}} (San Jose,
  California, USA) \emph{(\bibinfo{series}{CHI '07})}.
  \bibinfo{publisher}{Association for Computing Machinery},
  \bibinfo{address}{New York, NY, USA}, \bibinfo{pages}{1395–1404}.
\newblock
\showISBNx{9781595935939}
\urldef\tempurl%
\url{https://doi.org/10.1145/1240624.1240836}
\showDOI{\tempurl}


\end{thebibliography}



\end{document}